\documentclass[prx,twocolumn,superscriptaddress,secnumarabic,balancelastpage,longbibliography]{revtex4-1}
\usepackage[utf8]{inputenc}
\usepackage{amssymb}
\usepackage{amsmath}
\usepackage{babel} 
\usepackage[dvipsnames]{xcolor}
\usepackage{chngpage}
\usepackage{lgrind}        
\usepackage{color}         
\usepackage{graphics}      
\usepackage[pdftex]{graphicx}      
\usepackage{longtable}     
\usepackage{epsf}          
\usepackage{bm}            
\usepackage{asymptote}     
\usepackage{thumbpdf}
\usepackage{verbatim}
\usepackage{url}
\usepackage{MnSymbol}
\usepackage{physics}
\usepackage[colorlinks=true]{hyperref} 
\usepackage{enumitem}	

\usepackage{float}
\usepackage{times}
\usepackage{color}
\usepackage{verbatim}
\usepackage{soul,xcolor}
\setstcolor{red}
\usepackage{colortbl}

\usepackage{tabularx}

\usepackage[normalem]{ulem} 

\newcommand{\outline}[1]{}   
\renewcommand{\hl}[1]{#1}
\makeatletter
\AtBeginDocument{
\g@addto@macro{\appendix}{\renewcommand{\p@subsection}{\@Alph\c@section}}
}
\makeatother

\begin{document}

\title{Collective many-body dynamics in a solid-state quantum sensor \\ controlled through nanoscale magnetic gradients}
\affiliation{Department of Physics, Harvard University, Cambridge, Massachusetts 02138, USA}
\affiliation{School of Engineering and Applied Sciences, Harvard University, Cambridge, Massachusetts 02138, USA}
\affiliation{Materials Department, University of California, Santa Barbara, CA 93106, USA}
\affiliation{Harvard Quantum Initiative, Harvard University, Cambridge, MA 02138, USA}
\affiliation{Marian Smoluchowski Institute of Physics, Jagiellonian University in Krak\'ow, 30-348 Krak\'ow, Poland}
\affiliation{Department of Physics, University of California, Santa Barbara, CA 93106, USA}
\affiliation{Department of Chemistry and Chemical Biology, Harvard University, Cambridge, Massachusetts 02138, USA}

\author{Piotr Put$^1$}
\thanks{These authors contributed equally to this work}
\author{Nathaniel T. Leitao$^1$}
\thanks{These authors contributed equally to this work}
\author{Haoyang Gao$^{1}$}
\thanks{These authors contributed equally to this work}
\author{Christina Spaegele$^2$}
\thanks{These authors contributed equally to this work}
\author{Oksana Makarova$^{1,2}$}
\author{Lillian B. Hughes Wyatt$^{3}$}
\author{Andrew C. Maccabe$^{4}$}
\author{Matthew Mammen$^{4}$}
\author{Bartholomeus Machielse$^{1}$}
\author{Hengyun Zhou$^{1}$}
\author{Szymon Pustelny$^{1,5}$}
\author{Ania C. Bleszynski Jayich$^{6}$}
\author{Federico Capasso$^{2}$}
\author{Leigh S. Martin$^1$}
\author{Hongkun Park$^{1,7}$}
\author{Mikhail D. Lukin$^{1}$}

\begin{abstract}
Coherent collective dynamics of strongly interacting qubits are a central resource in quantum information science, with applications from quantum computing and simulation to metrology. While electronic spins interact strongly via dipolar couplings in dense solid-state ensembles, imperfections and positional disorder pose major obstacles to coherent correlated behavior, limiting their usefulness. Here, we \hl{realize}
collective many-body dynamics by combining time-dependent magnetic field gradients with global coherent control of dense electron spin ensembles in diamond. We control and probe the dynamics of nanometer-scale spin spirals, and, by exploiting Hamiltonian engineering that enhances the microscopic symmetry of the interactions, we observe a disorder-resilient collective spin evolution. Our results establish a pathway to interaction-enhanced quantum metrology and nanoscale imaging of materials and biological systems under ambient conditions.
\end{abstract}

\maketitle

\section{Introduction}
Coherent collective dynamics — where an isolated quantum many-body system evolves in a correlated, phase-coherent fashion — emerge either in systems composed from individually manipulated strongly coupled qubits, such as in quantum computers \cite{ebadi_quantum_2021,andersen_thermalization_2025,manovitz_quantum_2025},  or in ensembles with coherently controlled interactions \cite{choi_observation_2017, bornet_scalable_2023}.  
The latter do not typically involve individual particle control, but the resulting coherent dynamics could still be useful, e.g., for reducing quantum projection noise below the standard quantum limit \cite{kitagawa_squeezed_1993} or for amplifying weak signals amidst noisy readout \cite{davis_approaching_2016}. First proposed in the context of spectroscopy and atomic clocks \cite{wineland_squeezed_1994}, such dynamics have since been explored in a broad range of experimental quantum platforms, including Bose–Einstein condensates \cite{sorensen_many-particle_2001, gross_nonlinear_2010, riedel_atom-chip-based_2010}, optical-cavity-mediated atomic ensembles \cite{schleier-smith_squeezing_2010, hosten_measurement_2016, colombo_time-reversal-based_2022, li_improving_2023}, and trapped ions interacting through phonons \cite{bohnet_quantum_2016, franke_quantum-enhanced_2023}, utilizing effective all-to-all couplings  mediated \textit{indirectly} by a bosonic mode. Recently, strongly interacting systems based on \textit{direct} dipolar spin interactions have been realized. Due to the angular averaging and sensitive position dependence of dipolar couplings, collective dynamics required the use of two-dimensional systems, such as ordered lattices of neutral atoms \cite{bornet_scalable_2023}, or itinerant ultracold molecules \cite{li_tunable_2023, miller_two-axis_2024} and neutral atoms \cite{douglas_spin_2024}. More recently this approach has been extended to solid-state systems, which are particularly promising for realizing novel quantum sensing modalities such as nanoscale imaging in biological and material science \cite{aslam_quantum_2023, zhou_robust_2023, arunkumar_quantum_2023,arai_fourier_2015, grinolds_subnanometre_2014}. 
While the dipolar interactions in such systems have recently been harnessed to amplify small signals \cite{gao_signal_2025} and to generate spin squeezing in sub-ensembles of spin defects \cite{wu_spin_2025}, their utility is severely limited by positional disorder inherent to solid-state spin ensembles. This lead to large variation in local coupling strengths, inducing fast dephasing \cite{kwasigroch_synchronization_2017, gao_signal_2025}, thereby preventing the realization of large-scale collective dynamics.

In this Article, we \hl{realize} coherent, collective, \hl{many-body} dynamics in a solid-state nanoscale sensor hosting positionally disordered spins in a three-dimensional sample. This is achieved by integrating local control through magnetic field gradients with symmetry-engineering of microscopic interactions through Floquet pulse sequences. Specifically, we overcome the effects of angular averaging of the dipolar system by initializing a spatially inhomogeneous, spiral-like state with a strong magnetic field gradient, and further achieve disorder-robust collective \hl{nonlinear} dynamics by engineering an SU(2)-symmetric interaction that suppresses previously observed disorder-induced  relaxation \cite{gao_signal_2025, wu_spin_2025}. Using nanoscale imaging, we directly probe these underlying physical mechanisms, revealing collective, coherent spin exchange driving the evolution of spin spirals. \hl{This observation of collective one-axis-twisting like dynamics in a solid-state system 
 opens up opportunities for substantial metrological gain in practical, nanoscale quantum sensing.}

\section{Controlling dense spin ensembles in strong magnetic gradients}
\begin{figure}[btp!]
\begin{center}
\includegraphics[width=88mm]{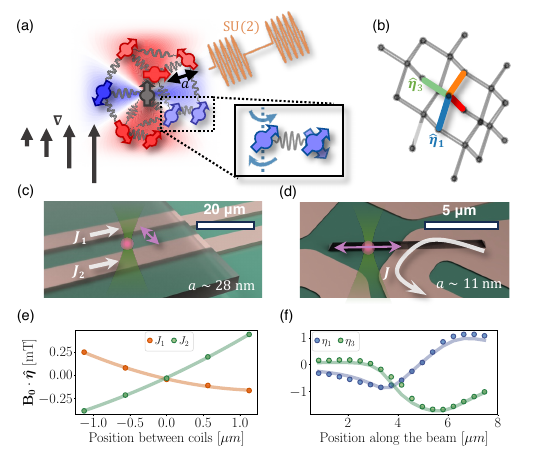}
\caption{Overview of the experimental platform. 
        (a) Schematic illustration of the experimental system, showing a positionally disordered three-dimensional ensemble of spins interacting via anisotropic magnetic dipolar coupling (red and blue coloring) and subject to a global MW field, which engineers interactions between spins, and a magnetic field gradient $\bm{\nabla}$. Positional disorder of the spins generically leads to fast dephasing driven by closely coupled pairs (inset), preventing collective dynamics.
        (b) The presence of four crystallographic groups of NV centers, $\bm{\eta_{i}}$, that also leads to different effective gradient directions $\bm{\nabla_{i}}$.
        (c) Illustration of the bulk sample device, showing a diamond plate placed atop a two-wire chip that generates a magnetic field gradient tunable via the ratio of currents on the two wires. Green laser illumination is used for spin initialization, red fluorescence indicates spin-state-dependent readout from a confocal spot. 
        (d) Illustration of the nanobeam device, showing a piece of a diamond (black beam) with a dense ensemble of NV centers placed atop a microcoil that generates an inhomogeneous magnetic field $\bm B_{0}$. 
        (e) Local magnetic field extracted from ESR measurements for the bulk sample, taken between microcoil wires (pink arrow in C) for currents running in each wire. Solid lines show simple quadratic fits used to determine the magnetic field gradient. 
        (f) ESR measurements along the nanobeam (pink arrow in D) for two different NV groups. Solid lines show finite-element simulations of the spatially varying magnetic field, projected onto the quantization axes of NV groups 1 and 3. The measurements were performed at one-tenth of the maximum gradient strength used in this work.}
\label{fig: 1}
\end{center}
\end{figure}

Our experimental system, illustrated in Fig.~\ref{fig: 1}(a), consists of a dense ensemble of nitrogen-vacancy (NV) spins in a diamond crystal~\cite{doherty_nitrogen-vacancy_2013}, interacting via magnetic dipole-dipole interactions. The NV electronic spins are initialized and read out using a green laser, and coherently manipulated via a globally applied microwave field, all under ambient conditions. During the experiment, we polarize a small region within the diamond, containing between $10^3$-$10^4$ spins. In this work we use two samples: a bulk diamond plate ($\sim 0.25$~ppm NV density) placed on top of two parallel gold wires (Fig.~\ref{fig: 1}(c)) and a denser NV sample ($\sim 3.8$~ppm) shaped into a triangular nanobeam (300 nm in size) and positioned atop a wire with a narrow constriction (Fig.~\ref{fig: 1}(d)).

Pulsed electric currents applied to the microcoils generate a spatially varying (inhomogeneous) magnetic field, $\bm{B}_{0}$ (see Fig.~\ref{fig: 1}(a)), which enables local spin control and spatially resolved readout. NV centers can be oriented along four different symmetry axes of the host diamond lattice (Fig.~\ref{fig: 1}(b)) and each group can be addressed individually by applying an external magnetic field aligned along it, such that each NV group experiences a different effective magnetic field gradient,
\begin{equation}
    \bm{\nabla}=  \nabla \left( \bm B_0 \cdot  \hat{\bm\eta}\right),
\end{equation}
where $\hat{\bm\eta}$ denotes a unit vector along the crystallographic symmetry axis of the NV center group. The gradient direction can be continuously tuned by changing the combination of currents running through microwires (Fig.~\ref{fig: 1}(e)), affecting the local magnetic field $\bm B_0$. Additionally, choosing a different NV group (Fig.~\ref{fig: 1}(f)) also discretely controls the gradient direction. 

To characterize the spatial structure and strength of the control fields, we measure the electron spin resonance (ESR) frequency for a bulk diamond device (Fig.~\ref{fig: 1}(e)) and the nanobeam (Fig.~\ref{fig: 1}(f)). We observe a clear spatial variation in the projected magnetic field $\bm{B}_0 \cdot \hat{\bm{\eta}}$, which agrees well with the predicted field profile. The resulting magnetic field gradients can reach up to $\sim 2.2$ mT/\textmu m, exceeding interaction energy at the typical spin–spin spacing by at least an order of magnitude. This enables the creation of spatially structured spin patterns (spin textures) within experimental coherence time and nanoscale resolution imaging required for investigating their dynamics (Methods).

\section{Probing dynamics of nanoscale spin spirals}
\begin{figure*}[btp!]
\begin{center}
\includegraphics[width=180mm]{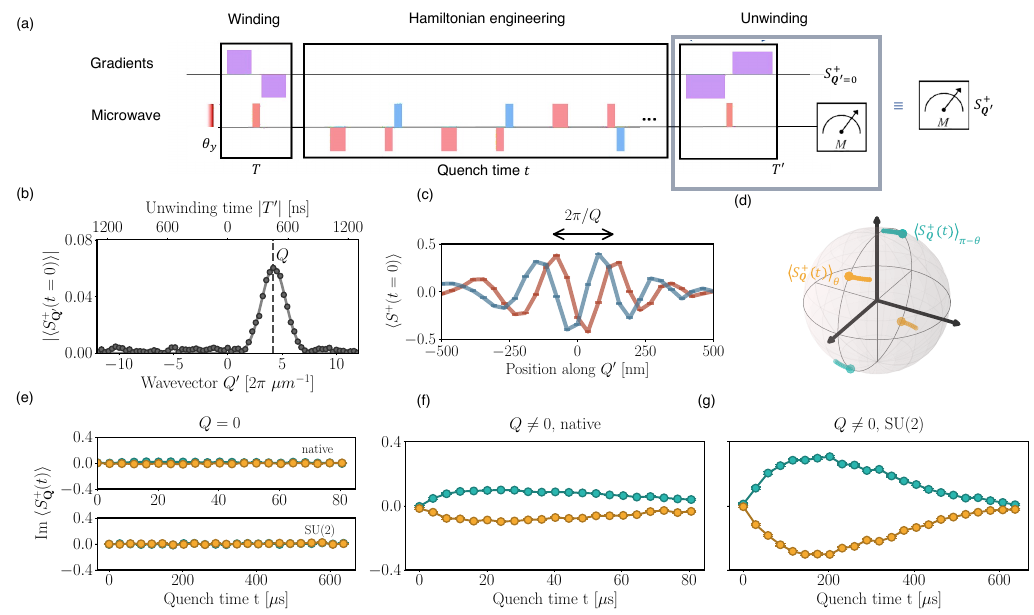}
\caption{Probing dynamics of conical spin spirals.
(a) Schematic of the measurement protocol, consisting of three stages: spin state preparation (\textit{winding}), quench under a Floquet-engineered Hamiltonian, and readout/imaging of the spin spiral in \textit{unwinding} stage.
(b) Benchmarking of the winding and unwinding protocol, omitting quench, demonstrating robust and reversible preparation and readout of spin spirals. Markers indicate measured points, solid line is a simple gaussian fit.
(c) Imaging of a prepared nanoscale spin texture. Measured $\hat{x}$ (red) and $\hat{y}$ (blue) components of the spin spiral showing spin texture wound at the target pitch $\bm Q$.
(d) Initial spin states used to probe many-body dynamics, shown on the Bloch sphere. Preparation of antipodal states (pairwise colored) enables cancellation of spurious global rotations, isolating the genuine many-body dynamics of the spin spiral \cite{gao_signal_2025}.
(e) Measurement of many-body dynamics when no spiral is wound, reported in unit of Bloch sphere radius. Top panel contains data for a native dipolar interaction, while the bottom panel corresponds to the engineered SU(2) Hamiltonian. Without a spiral no time-dependent signal is observed. Data is taken with the bulk sample device throughout the paper, unless explicitly stated otherwise.
(f) Measurement of the spin spiral precession dynamics for a native dipolar Hamiltonian.
(g) Measurement of the spin spiral dynamics for a Floquet engineered SU(2) Hamiltonian taken over longer quench timescale.
}
\label{fig: 2}
\end{center}
\end{figure*}

Utilizing these strong gradient fields, we investigate dynamics of the spatially inhomogeneous spin texture with the measurement sequence illustrated in Fig.~\ref{fig: 2}(a). By simultaneously controlling local magnetic gradients and global microwave pulses, the experiment is divided into three stages.

In the first stage, referred to as \textit{winding}, spins are optically polarized along the $\hat{z}$ direction set by the crystallographic quantization axis and then globally rotated by an angle $\theta$ about the $\hat{y}$-axis using a microwave pulse. A magnetic field gradient is subsequently applied, imprinting a spatially varying Zeeman shift $\bm \nabla \cdot \bm r_j\, S_j^z$ across the ensemble. This phase imprinting is done in two equal duration blocks separated by a microwave $\pi$-pulse to decouple magnetic disorder.
The resulting state is a spatially modulated spin configuration resembling a conical spin spiral~\cite{jepsen_long-lived_2022} winded at the cone angle $\theta$. The magnitude of the spiral wavevector $\bm{Q}=T \bm{\nabla}$ can be controlled by adjusting the winding time $T$ (see Methods and Supplementary Information (SI)).
In the second stage, called \textit{quenching}, we apply periodic microwave driving to engineer an effective many-body Hamiltonian via Floquet engineering~\cite{martin_controlling_2023}. The quench duration $t$ is controlled by varying the number of Floquet cycles. 
In the final stage, \textit{unwinding}, we apply a reversed gradient for a variable time $T'$, corresponding to a measurement wavevector $\bm{Q'}$. By measuring the global spin signal after this step, we effectively measure the Fourier mode $S^+_{\bm Q'} = \sum_j e^{-i \bm Q' \cdot \bm r_j} S_j^+$, where $S_j^+ = S_j^x + i S_j^y$ represents the spin coherence at site $j$.

To validate this winding–unwinding protocol on the bulk sample device, we prepare a spin spiral with cone angle $\theta = \pi/2$ and a pitch $Q = 2\pi/(0.242\,${\textmu m}$)$ and immediately reverse the process without any intermediate quench. Varying the unwinding time (and thus $\bm Q'$), we measure the ensemble-averaged $x$ and $y$ spin components (Fig.~\ref{fig: 2}(b)). We observe a clear revival in the coherence magnitude $|\langle S^+_{\bm Q'}(t=0)\rangle|$ centered at $Q' = Q$, providing evidence of reversible dynamics under the gradient field. The finite width of this revival profile reflects the spatial extent of the spin-polarized region ($\Delta Q' \approx 2\pi / (0.4\,${\textmu m})). 

This data can be processed via nanoscale Fourier Magnetic Imaging (FMI)\cite{arai_fourier_2015} to reconstruct spin distributions from the inverse Fourier transform of $\langle S^+_{\bm Q'} \rangle$. The resulting spin spiral polarization is plotted in Fig.~\ref{fig: 2}(c) and matches the expected spatial profile imposed by the wavevector $\bm Q$, confirming our ability to both prepare and measure Fourier modes of spin polarization with high fidelity.

We next proceed to investigate the many-body dynamics of spin spirals on the bulk sample device,  focusing on the dynamics of the initialized Fourier mode $\boldsymbol{Q'}=\boldsymbol{Q}$.  To isolate intrinsic many-body dynamics from accumulated microwave pulse errors, we prepare four spin spirals based on two pairs of antipodal spin-coherent states, prepared prior to winding in the XZ plane (Fig.~\ref{fig: 2}(d)). This approach allows us to average out systematic errors and spurious rotations corrupting the intrinsic nonlinearity of the many-body evolution \cite{gao_signal_2025}.

In Fig.~\ref{fig: 2}(e), (f), (g), we show the extracted spiral dynamics by plotting the imaginary component of the Fourier amplitude, $\text{Im}\langle S^+_{\bm Q}(t) \rangle$, i.e. total $\hat{y}$ coherence measured \textit{after} unwinding, averaged across antipodes, as a function of quench time $t$. When no spiral is prepared ($Q = 0$, Fig.~\ref{fig: 2}(e)), no time-dependent signal is observed. For the native dipolar Hamiltonian (top panel) probed via an XY-16-type decoupling sequence (see SI), this arises from angular averaging of the dipolar anisotropy, which suppresses the mean field~\cite{gao_signal_2025, wu_spin_2025}. A similar absence of dynamics occurs for the Floquet-engineered, SU(2) symmetric Hamiltonian (bottom panel) probed via a DROID-type sequence (see SI), where the total spin of the globally polarized initial state is conserved preventing time evolution. 

In contrast, when a finite-wavelength spin spiral is initialized ($Q = 2\pi/(0.242~${\textmu m}) $\neq 0$), we observe clear dynamical evolution of the $\hat{y}$ coherence, indicating nonlinear interaction-driven many-body dynamics (Fig.~\ref{fig: 2}(f)). The second pair of antipodal initial states exhibits similar precession dynamics but with an opposite sign. For an SU(2) symmetric Hamiltonian (Fig.~\ref{fig: 2}(g)) we again observe a clear dynamical signal, similar to the one in Fig.~\ref{fig: 2}(g) but on a longer timescale and with significantly higher amplitude.

\section{Engineering dipolar spin exchange}
\begin{figure}[btp!]
\begin{center}
\includegraphics[width=88mm]{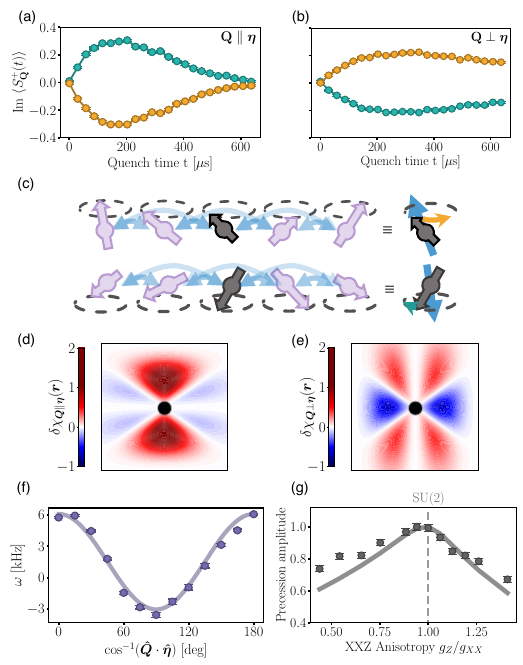}
\caption{Tuning dipolar spin dynamics via geometric and microscopic Hamiltonian anisotropy.
(a) Measurement of spin spiral dynamics for $\bm{Q} \parallel \bm{\eta}$.
(b) Opposite-sign, slower spiral precession measured for $\bm{Q} \perp \bm{\eta}$. 
(c) Semiclassical mechanism generating spiral precession due to exchange mean fields at the SU(2) point. The central spin (black) experiences an effective mean field from nearby spins. The transverse components of this field lead to spin precession. (Bottom panel) For a spiral starting from an initial state on the lower hemisphere of the Bloch sphere, the direction of the mean field and the sign of the precession are reversed.
(d) Illustration of the dipolar interaction, spatially modulated at the pitch of the spiral for a simplified case where quantization axis is parallel to gradient direction, $\delta \chi_{\bm Q \parallel \bm \eta}(\bm r)$.
(e)  Corresponding spatial modulation of the dipolar interaction for the case where the quantization axis is perpendicular to the gradient direction, $\delta \chi_{\bm Q \perp \bm \eta}(\bm r)$, sourcing an exchange field of opposite sign.
(f) Measured precession frequency as a function of geometric anisotropy $\boldsymbol{\hat{Q}} \cdot \boldsymbol{\hat{\eta}}$ for $Q = 2\pi/(0.242~${\textmu m}), showing tunable strength and sign of the exchange field. Solid line is the dipolar anisotropy $\mathcal{A}_{\hat{\bm\eta}}(\hat{\bm Q})\propto 3(\hat{\bm Q}\cdot \hat{\bm \eta})^2-1$ theoretically expected, see SI. 
(g) Normalized precession amplitude as a function of interaction anisotropy in the engineered XXZ Hamiltonian. Experimental data are shown as markers; solid line represents theory prediction.
}
\label{fig: 3}
\end{center}
\end{figure}

To investigate the microscopic mechanism responsible for the observed nonlinear spin spiral dynamics (Fig.~\ref{fig: 2}(f),~(g)), we take advantage of the tunability of the gradient direction. Specifically, we complement the earlier measurement performed with a spiral wavevector aligned to the NV quantization axis $\bm{Q} \parallel \bm{\eta}$, with a measurement where the spiral direction is perpendicular to the quantization axis $\bm{Q} \perp \bm{\eta}$. As shown in Fig.~\ref{fig: 3}(a),~(b), the two measurements exhibit qualitatively similar dynamics but with opposite signs of precession. This reversal provides a key signature of the underlying dipolar interaction mechanism. 

To understand these observations, we note that the NV spins interact through strong magnetic dipolar couplings, which, under the Floquet-engineered sequence used above, are described by a spin-exchange Hamiltonian with a global SU(2) spin rotation symmetry,
\begin{equation}
    H_{0}= \sum_{i<j} \frac{\mathcal{A}_{\hat{\bm \eta}}(\hat{\bm r}_{ij})}{r_{ij}^3}  \bm{S}_i \cdot \bm {S}_j ,
\end{equation}
providing a long-ranged, dipolar version of the quantum Heisenberg model \cite{leitao_scalable_nodate}.
In particular, the spatially anisotropic couplings  $\mathcal{A}_{\hat{\bm{\eta}}}(\hat{\bm r})\propto   3\left( \hat{\bm \eta}\cdot \hat{\bm r} \right)^2-1 $, can lead to either ferromagnetic or anti-ferromagnetic interactions depending on the orientation of the interacting spin pair relative to their mutual quantization axis $\hat{\bm \eta}$. 

We consider the mean-field evolution of a central spin $\bm S$ polarized in the $XZ$ plane of the Bloch sphere (black arrow and dot in Fig.~\ref{fig: 3}(c),~(d),~(e)). The mean field sourced at this spin is given by $\bm B_{\bm Q}=\sum_j \tfrac{\mathcal{A}_{\hat{\bm \eta}}(\hat{\bm r}_j)}{r_j^3}\langle \bm{S}_j \rangle_{\bm Q}$, where $\langle\bm{S}_j \rangle_{\bm Q}$ is the polarization of the $j$-th spin in spiral texture with wavevector $\bm Q$. For an ideal conical spiral (see SI), this becomes $\bm{B}_{\bm Q}= \big(\sum_j \delta\chi_{\bm Q}(\bm r_j)\big)\bm S_\perp + \big(\sum_j \tfrac{\mathcal{A}_{\hat{\bm \eta}}(\hat{\bm r}_j)}{r_j^3}\big)\bm S$, with $\delta\chi_{\bm Q}(\bm r)= \tfrac{\mathcal{A}_{\hat{\bm \eta}}(\hat{\bm r})}{r^3}\tfrac{1-\cos(\bm Q\cdot\bm r)}{2}$ being the \textit{transverse} field contribution from a spin at $\bm r$, see Fig.~\ref{fig: 3}(c). Integrating over the ensemble yields the effective \textit{exchange field strength}
\begin{align}\label{eq:exchange-field}
\chi_{\bm Q} &= \int d\bm r\, \rho(\bm r)\, \delta \chi_{\bm Q}(\bm r) 
= \int d\bm r\, \rho(\bm r)\, \frac{\mathcal{A}_{\hat{\bm \eta}}(\hat{\bm r})}{r^3}\,\frac{1-\cos(\bm Q\cdot \bm r)}{2},
\end{align}
where $\rho(\bm r)$ is the polarization density.

Intuitively, the exchange field reflects dipolar coupling modulated at the spiral pitch (Fig.~\ref{fig: 3}(d),~(e)). For $\bm Q=0$ it vanishes due to collinear spin alignment. When $\bm Q \parallel \hat{\bm \eta}$, anti-ferromagnetically coupled spins (red regions in Fig.~\ref{fig: 3}(d)) dominate and generate a positive torque, while for $\bm Q \perp \hat{\bm \eta}$ ferromagnetically coupled spins (blue regions in Fig.~\ref{fig: 3}(e)) dominate, yielding a negative torque consistent with reversed sign of the spiral precession (Fig.~\ref{fig: 3}(b)).  This experimental tunability is borne out by continuously rotating the spiral wavevector, where we experimentally observe that the spiral precession rate $\omega$, extracted from the early-time precession angle,  follows the dipolar anisotropy (Fig.~\ref{fig: 3}(f)). By modulating dipolar anisotropy and tuning geometry, we can generate tunable mean-field leading to nonlinear spin dynamics in stark contrast to longitudinal Ising-type fields~\cite{block_scalable_2024, gao_signal_2025, wu_spin_2025}, which average out in three-dimensional dipolar systems.

Comparing the spin dynamics in Fig.~\ref{fig: 2}(f),~(g), we find that evolution under the Floquet-engineered SU(2) Hamiltonian is markedly slower than under the native interaction. This slowdown results from Floquet engineering rescaling the exchange term of the Hamiltonian \cite{martin_controlling_2023}—which drives spin-spiral dynamics—by a factor of 1/3. Remarkably however, despite the slower early-time precession, the spiral ultimately develops a much larger precession amplitude, prompting us to probe the role of spin-rotational SU(2) symmetry by continuously tuning the anisotropy of the effective Hamiltonian.

Adjusting the spacing between microwave pulses in Floquet engineering~\cite{martin_controlling_2023}, we realize a family of U(1)-symmetric XXZ Hamiltonians with tunable relative interaction strengths: $\bm{S}_i \cdot \bm{S}_j \rightarrow g_{XX}(S_i^x S_j^x + S_i^y S_j^y) + g_Z S_i^z S_j^z $. We measure the maximum amplitude of spiral precession as a function of the anisotropy ratio $g_Z/g_{XX}$ (Fig.~\ref{fig: 3}(g)), and observe a maximum at the SU(2)-symmetric point ($g_Z/g_{XX} = 1$). This observation is explained by the accelerated dephasing further away from the SU(2) point (see Fig.~S3(a),~(b)), where the transverse spin components are no longer globally conserved \cite{jepsen_transverse_2021}.

\section{Microscopic polarization driving spiral dynamics}
\begin{figure*}[btp!]
\begin{center}
\includegraphics[width=180mm]{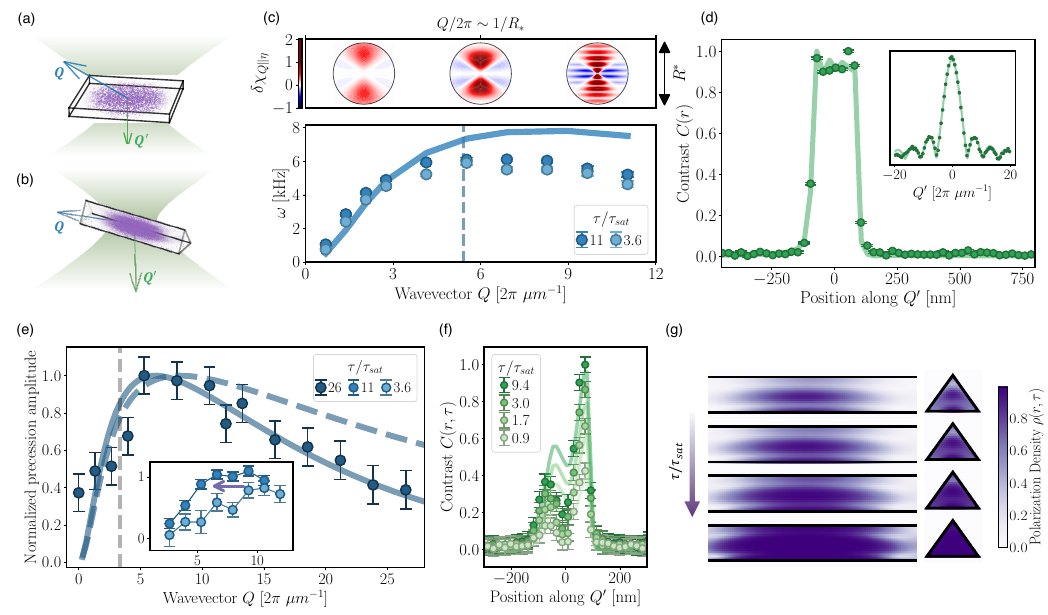}
\caption{Magnetic imaging of microscopic polarization driving spiral dynamics. 
    (a) Spin polarization geometry in bulk diamond: disc-shaped region set by NV layer thickness (axial) and optical pumping (transverse). Gradient directions for data in (C) (blue) and imaging in (D) (green) are indicated. 
    (b) Analogous geometry in the nanobeam device. 
    (c) (Top) Origin of non-monotonic precession amplitude: saturation wavevector $Q_{*}$ is inversely related to polarization extent $R_{*}$. (Bottom) Measured spiral precession frequency $\omega$ versus wavevector $Q$ in bulk diamond. Vertical line: $Q = 2\pi/(0.185~${\textmu m}). Solid lines are theory predictions.   
    (d) FMI taken across the bulk sample, used to extract NV layer thickness (185~nm). Inset: raw data versus $Q'$, with fit assuming a rectangular profile. 
    (e) Maximum normalized precession amplitude measured versus $Q$ for NV group~1 in the nanobeam device. Vertical dashed line: $Q = 2\pi/(0.3~${\textmu m}). Theory: dashed line—ideal spiral; solid line—with measured spiral winding loss (see SI). (Inset) Amplitude measured versus $Q$ for shorter optical pumping times, showing saturation shift consistent with larger polarization extent. 
    (f) FMI along the nanobeam short axis, revealing nanoscale polarization variations from nanophotonic interference.  Experimental data are shown as markers; solid line represents nanophotonics model predictions.
    (g) Theory cross-sections of NV polarization in the nanobeam for increasing pump times. Spatial structure arises from optical interference of the green pump light.
}

\label{fig: 4}
\end{center}
\end{figure*}

The spin-exchange mechanism described above suggests that many-body dynamics can be tuned through the spiral pitch $\bm Q$ (see Eq.~\ref{eq:exchange-field}). After exploring the role of wavevector direction, we now vary its magnitude. Specifically, we extract the early-time spiral precession frequency $\omega$ as a function of wavevector magnitude (Fig.~\ref{fig: 4}(c), bottom panel).
The precession rate rises rapidly, then saturates and decays at larger wavevectors. To identify the characteristic length scale at which saturation occurs, we exploit a different imaging axis $\hat{\bm{Q}}'$ (see green arrows in Fig.~\ref{fig: 4}(a),~(b)). From the polarization profile (Fig.~\ref{fig: 4}(d)), we observe NVs confined axially by the sample thickness (185 nm), consistent with a fit to raw data in the Fourier space (inset). We note here that the saturation in dynamics occurs around  $Q_{*} = 2\pi/(0.185~${\textmu m}) (dashed line in Fig.~\ref{fig: 4}(c)), indicating sensitivity to the extent of polarized region in the sample.  

To corroborate these results, we study spiral dynamics on the nanobeam device. Using the first NV group, we measure precession amplitude versus $Q$ (Fig.~\ref{fig: 4}(e)), which in this case probes the exchange field strength (Eq.~\ref{eq:exchange-field}, SI). The amplitude shows  initial increase, saturation, and decay, with saturation occurring again at the scale of the polarized region (300 nm width of the beam). Notably, the saturation wavevector depends on the optical pumping duration $\tau$ (inset of Fig.~\ref{fig: 4}(e), pumping time normalized by the calibrated saturation time $\tau_{sat}$, see SI), in contrast to the results from a bulk sample device (dark and light blue points in Fig.~\ref{fig: 4}(c)). We again elucidate these observations using FMI. Taking an image perpendicular to the nanobeam axis (Fig.~\ref{fig: 4}(f)), we reveal a dip in the polarization profile arising from nanophotonic interference of the pump light, confirmed by electric field simulations (Methods). Simulated NV polarization (Fig.~\ref{fig: 4}(g)) shows that short pumping creates two lobes, which merge into a uniform profile with longer pumping, changing the extent of microscopic polarization and further pointing to the sensitivity of the many-body dynamics to the system size.

To understand this feature of dipolar many-body dynamics, we analyze $\chi_{\bm Q}$ (Eq.~\ref{eq:exchange-field}) for an isotropic polarization radius $R_*$ (Fig.~\ref{fig: 4}(c), top panel). For $Q/2\pi < 1/R_*$, spins align with the central spin, driving negligible exchange. With increasing $Q$, distant spins contribute strongly via dipolar interactions, and spherical shells, of thickness $dr$ at radius $r$, balance density $4\pi r^2dr$ against the $1/r^3$ decay, maximizing exchange near $Q/2\pi \sim 1/R_*$. At larger $Q$, dipolar anisotropy reduces the exchange, causing the observed decay \cite{warren_generation_1993}. 

\section{Coherent, collective dynamics in a disordered dipolar spin system}
\begin{figure*}[btp!]
\begin{center}
\includegraphics[width=180mm]{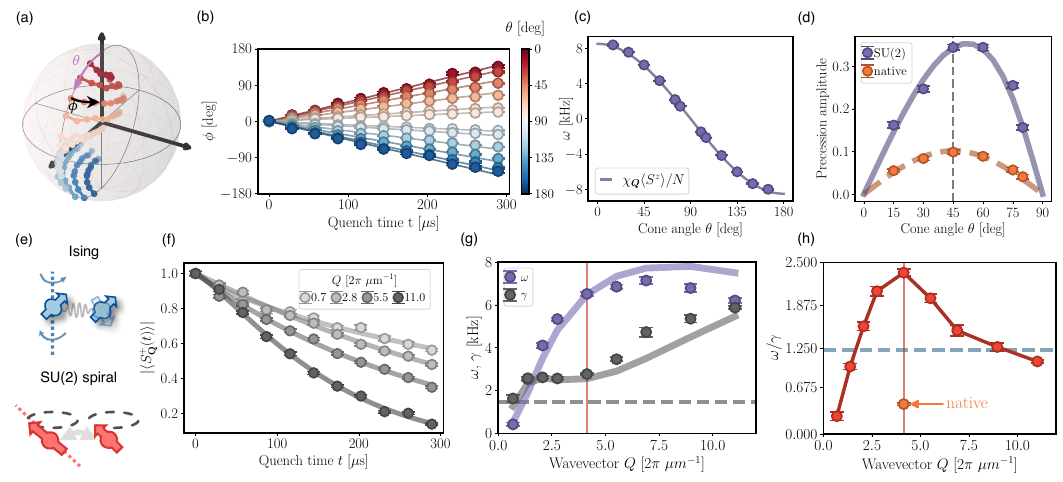}
\caption{Collective nonlinearity in a disordered, three-dimensional dipolar spin system.
    (a) Measured dynamics of spin spirals prepared at cone angle $\theta$ and precessing by angle $\phi$. Most transparent point marks maximum quench time $t=289.44$~{\textmu}s. Data used in b–d was collected with improved drive/gradient homogeneity at $Q = 2\pi/(0.242~${\textmu m}), shown by the thin red line in (g),(h).  
    (b) Measured spiral precession angle $\phi$ versus quench time for a range of initial cone angles $\theta$. Solid lines are linear fits used to extract $\omega$. 
    (c) Extracted precession rate versus cone angle $\theta$. The fitted cosine dependence (solid line) is characteristic of OAT dynamics.  
    (d) Measurement of precession amplitude versus cone angle $\theta$ (markers). Purple line: numerical simulations multiplied by measured decoherence (Fig.~S1(b)). Dashed orange: naive model 
    $\propto\sin(\theta)\cos(\theta)$. 
    (e) Illustration of a pair of strongly coupled spins. In the Ising case, local interactions and quantum fluctuations lead to fluctuating $\hat{z}$-field (vertical dashed line), causing random precession (curved arrows) and fast dephasing, whereas for a long-wavelength spiral evolving under an SU(2) Hamiltonian, nearby spins contribute a field nearly aligned (red dashed line) with the target spin, preventing local dephasing. 
    (f) Decay curves for SU(2) conical spiral with $\theta = 45^{\circ}$ at varying wavevectors. Solid lines are fits to stretched exponential decay.
    (g) Precession frequency $\omega$ and decay rate $\gamma$ measured as functions of the spiral wavevector $Q$ for SU(2) interaction. The horizontal dashed line marks the extrinsic decay rate; solid lines denote theoretical predictions.  
    (h) Quality factor $\omega/\gamma$ computed from (g). The horizontal dashed line indicates the theoretical maximum for a two-dimensional Ising model of dilute, positionally-disordered spins with unity spin polarization, while an orange marker indicates the measured quality factor for native interactions.  
}

\label{fig: 5}
\end{center}
\end{figure*}

We next explore the long-time dynamics of the spin spirals on the bulk sample device. Specifically, we prepare spin spirals with  different cone angles $\theta$ (Fig.~\ref{fig: 5}(a)) and monitor their evolution at $Q = 2\pi/(0.242~${\textmu m}), where the spin-exchange field is strongest (Fig.~\ref{fig: 4}(c)). Experimentally, we observe a linear growth of the precession angle $\phi$ (Fig.~\ref{fig: 5}(b)), with a rate that exhibits a cosine dependence upon the cone angle $\theta$, consistent with the  $S_z$-dependent precession frequency 
(Fig.~\ref{fig: 5}(c)). These observations closely resemble the nonlinear dynamics generated by the one-axis twisting (OAT) Hamiltonian~\cite{kitagawa_squeezed_1993} for $N$ spins, $H_{\text{OAT}}=\chi \left(S^z\right)^2/N$, with an effective twisting rate $\chi=\chi_{\bm Q} = 8.51(2)$~kHz.  Fig.~\ref{fig: 5}(d) shows the maximum spiral twisting amplitude for different interaction types. SU(2) engineered interactions (purple points) produce a large twisting — considerably exceeding values resulting from both the native interaction Hamiltonian (orange points) as well as those measured previously in similar systems
\cite{gao_signal_2025,wu_spin_2025} — and a pronounced asymmetry toward larger cone angles that reflects the curvature of the collective Bloch sphere.

To understand these observations, we note that in the recent studies of two-dimensional disordered systems \cite{gao_signal_2025, wu_spin_2025}, twisting dynamics were found to be limited by strongly interacting spin pairs which induce fast dephasing (top panel, Fig.~\ref{fig: 5}(e)). To understand why such dephasing does not limit the twisting signal in our case, we note that nearby spin pairs in a long-wavelength spiral are approximately colinear (bottom panel, Fig.~\ref{fig: 5}(e)). Under an SU(2)-symmetric Hamiltonian, such colinear pairs do not contribute appreciably to the many-body dynamics, rendering the spiral dynamics insensitive to local dynamics in these strongly coupled pairs, or other microscopic details. More generally, the spin spirals employed here form slow, hydrodynamic modes of the many-body system under SU(2)-symmetric interactions \cite{krajnik_undular_2020, glorioso_hydrodynamics_2021}. These modes remain protected from relaxation up to the macroscopic spin-transport timescale, which is set by the tunable spiral wavelength $2\pi/Q$ and the transport universality class of the dipolar Heisenberg model\cite{leitao_scalable_nodate}. Experimentally, this protection manifests as a markedly slower decay of spiral amplitude for longer spiral wavelengths, as shown in Fig.~\ref{fig: 5}(f).

Furthermore, by comparing the fitted decay rate $\gamma$ with the precession frequency $\omega$ in Fig.~\ref{fig: 5}(g), we find a striking separation of scales. The maximum separation occurs when the spiral pitch is comparable to the inverse linear system size, consistent with theoretical predictions (solid lines in Fig.~\ref{fig: 5}(g)). In Fig.~\ref{fig: 5}(h), we explicitly plot the quality factor $\omega/\gamma$ versus spiral pitch and compare it with the analytic limit set by local dynamics in a dilute, positionally disordered two-dimensional Ising model, predicting maximum $\omega/\gamma =\sqrt{3/2}\sim 1.22$ (dashed line, see SI). While for the system with native interactions, quality factor is below this limit,  in the case of SU(2) interactions,  
for a broad range of wavevectors, our observations clearly exceed  this ideal limit, demonstrating collective nonlinear dynamics in a positionally disordered dipolar system.  

\section{Discussion and outlook}
Our experiments demonstrate that nanoscale magnetic field gradients \hl{induce emergent collective dynamics in disordered dipolar ensembles.} 
While magnetic gradients have been used previously to probe dipolar  interactions in  nuclear spin ensembles \cite{warren_generation_1993,romalis_transverse_2001, ledbetter_nonlinear_2002}, it is the combination with  Floquet-engineered SU(2)-symmetric interactions 
\cite{choi_robust_2020, martin_controlling_2023} that 
\hl{mitigates the effects of disorder, engineering 
coherent collective dynamics.} In particular, we have shown that the far-from-equilibrium relaxation of long-wavelength spin spirals under SU(2)-symmetric dipolar interactions \hl{generates} coherent OAT-like 
dynamics.

Twisting dynamics  with all-to-all Ising interactions provides the simplest, textbook example of collective nonlinearity from Ising interactions among $N$ spins, with a quality factor $\omega/\gamma = \mathcal{O}(\sqrt{N})$, capable of generating scalable amplification and spin squeezing \cite{kitagawa_squeezed_1993}. By contrast, generic short-range Ising interactions yield a quality factor that does not scale with system size, $\omega/\gamma = \mathcal{O}(1)$ \cite{foss-feig_entanglement_2016} (dashed line in Fig.~\ref{fig: 5}(h) for a dilute dipolar system, see SI). The use of spin spirals allows us to surpass this limit (Fig.~\ref{fig: 5}(h)) with substantially larger quality factor via SU(2)-symmetry engineering of dipolar interactions. 
Theoretical analysis (see \cite{leitao_scalable_nodate} and SI) predicts that in isolated, disordered dipolar systems, such spiral-mediated twisting can in principle achieve an ideal $\mathcal{O}(\sqrt{N})$ quality factor due to the separation of timescales between spiral relaxation and exchange-driven twisting. Specifically, for dipolar interactions in three dimensions, one expects the twisting rate $\omega$ to be independent of the system size - provided the spiral wavevector is tuned to the inverse system size, $Q_* =\mathcal{O}(1/R_*)= \mathcal{O}(1/N^{1/3})$ (red line in Fig.~\ref{fig: 5}(g),~(h)) - while the relaxation rate $\gamma$ is parametrically slower in larger systems due to emergent spin hydrodynamics \cite{leitao_scalable_nodate}, resulting in collective, scalable twisting dynamics directly from dipolar interactions.

Our results can be extended along several directions. Probing intrinsic spiral relaxation  and its connection to emergent spin hydrodynamics can provide new insights into the dynamics of such a complex system \cite{gopalakrishnan_kinetic_2019, ye_emergent_2020, zu_emergent_2021}. While related phenomena have been explored in electronic and nuclear spin-$1/2$ systems \cite{emsley_spin_2009, dikarov_direct_2016}, ultracold atoms~\cite{levy_collective_1984, jepsen_spin_2020, jepsen_transverse_2021, jepsen_long-lived_2022, hild_far--equilibrium_2014}, and spin-orbit-coupled semiconductors~\cite{bernevig_exact_2006, koralek_emergence_2009}, our platform offers distinct features, including tunable SU(2) symmetry, long-range interactions, magnetic frustration, and intrinsic positional disorder, allowing for spatially resolved studies of such complex quantum many-body dynamics \cite{acebron_kuramoto_2005, reimann_nonequilibration_2023, medenjak_isolated_2020}.

At the same time, the disorder-robust collective nonlinearity observed here can further be applied to realize interaction-enhanced metrology. With long coherence times \cite{gao_dressed-state_nodate} and high-fidelity readout \cite{arunkumar_quantum_2023}, entanglement generation and spin squeezing \cite{leitao_scalable_nodate} become feasible, while scalable signal amplification \cite{gao_signal_2025, leitao_optimally_nodate} promises substantial magnetic sensitivity gains. 
Exploiting long-wavelength spiral textures and Hamiltonian engineering, we introduced a mechanism to regulate positional disorder that extends beyond methods based on depolarizing strongly coupled spins \cite{wu_spin_2025},  thus preserving the density of quantum sensors and associated magnetic sensitivity \cite{zhou_robust_2023}. This is particularly relevant since the experimental platform introduced here is a prime candidate for a nanoscale magnetic resonance imaging~\cite{budakian_roadmap_2024} system, owing to its high spatial resolution (below 20 nm in this work)  inherited directly from the high spin density and strong gradients. Those versatile imaging capabilities can be further extended making use of multiplexed detection \cite{guo_wide-field_2024}, compressed sensing \cite{arai_fourier_2015, t_amawi_three-dimensional_2024}, and applied to measure spatiotemporal correlations \cite{arai_fourier_2015, koyluoglu_interaction-enhanced_nodate},  naturally complementing the intrinsic magnetic sensing modalities of dense NV ensembles \cite{zhou_robust_2023, arunkumar_quantum_2023}. Together, these results establish a versatile platform for nanoscale sensing and magnetic imaging, with quantum advantage within reach under ambient conditions.
\section{Methods}

\subsection{Gradient chip}
\subsubsection{Nanobeam device}
\begin{figure*}[btp!]
  \centering
  \includegraphics[width=180mm]{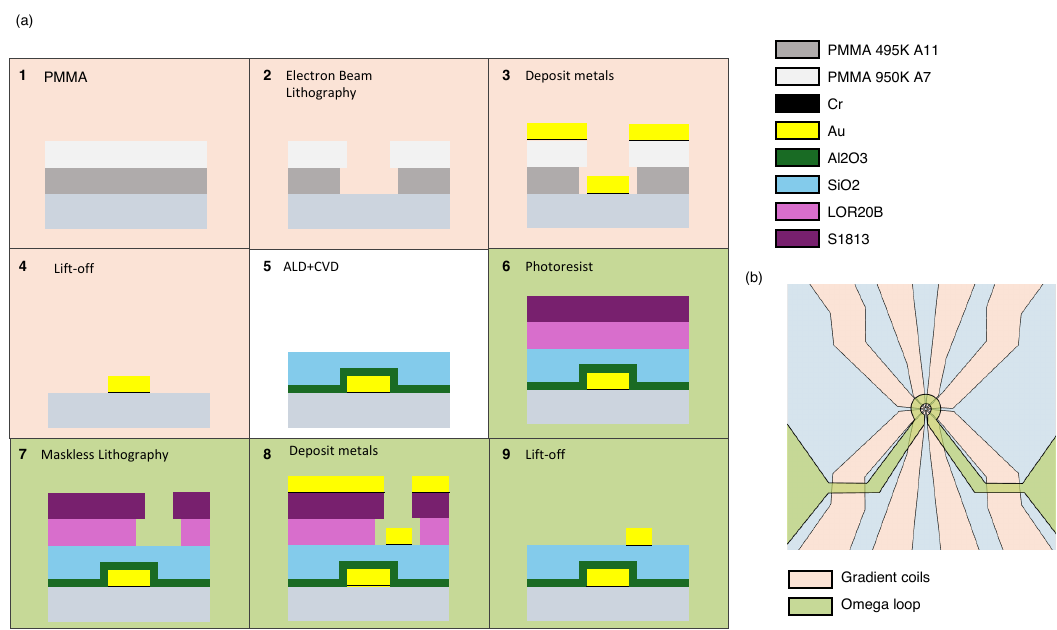}
  \caption{Gradient chip fabrication. (a) Two-layer gradient chip fabrication process divided into nine steps. Different material layers are indicated by different colors. (b) Design of the fabricated gradient chip. The two layers—containing the gradient coils and the omega loop used for microwave (MW) driving—are highlighted.}

\label{fig: SI1}
\end{figure*}

To generate strong magnetic field gradients and a microwave (MW) drive for the ensemble of dipolar spins in a diamond nanobeam, a custom, two-layer chip was fabricated following a series of steps:

\begin{itemize}
    \item Substrate preparation: Silicon carbide substrates (23 mm x 23 mm x 0.35 mm) were cleaned with acetone, isopropanol (IPA), piranha solution (H\textsubscript{2}SO\textsubscript{4}:H\textsubscript{2}O\textsubscript{2}, 4:1), buffered hydrofluoric acid (BHF, 5:1), and oxygen plasma (100~W, 40~sccm, 1~min) to remove organic residues, surface oxides, and enhance resist adhesion.

    \item Gradient coil fabrication: Two layers of polymethyl methacrylate (PMMA) resist (PMMA 495K followed by PMMA 950K) were spin-coated onto the substrate. A conductive espacer layer was deposited to mitigate charging during electron beam lithography (EBL).  Gradient coils were defined by EBL and developed in MIBK:IPA (1:3). A descum step was performed using oxygen plasma. A metal stack of chromium (Cr, 10~nm) and gold (Au, 500~nm) was deposited by electron beam evaporation. Liftoff was performed in Remover PG with mild ultrasonication.

    \item Dielectric spacer deposition: An alumina (Al\textsubscript{2}O\textsubscript{3}) layer (30~nm) was deposited by atomic layer deposition (ALD). A silicon dioxide (SiO\textsubscript{2}) layer (500~nm) was deposited by plasma-enhanced chemical vapor deposition (PECVD).

    \item $\Omega$-loop fabrication: Two layers of resist (LOR20B followed by S1813) were spin-coated onto the sample. The $\Omega$-loop was patterned using a maskless lithography aligner, aligned to features near the device center. After development, a Cr (10~nm) / Au (700~nm) metal stack was deposited by electron beam evaporation. Liftoff was performed in Remover PG.

    \item Contact pad opening: A photoresist (S1818) was spin-coated to define the bonding pad openings. Exposed dielectric layers were removed by buffered oxide etch (BOE). A final cleaning step was performed in Remover PG.
\end{itemize}
The fabrication steps, as well as the overall chip design, are shown in Fig.~\ref{fig: SI1}. The fabricated chip under white light and confocal microscope is shown in Fig.~\ref{fig: SI2}(a),~(b). The design of the chip allows for a switchable current to be applied through any of the four small constrictions, enabling a time-dependent magnetic field gradient Fig.~\ref{fig: SI2}(d),~(e). On the second layer of the chip, an $\Omega$-loop is patterned, which is used to coherently drive spins, allowing for disorder decoupling and Floquet engineering of a desired form of an interaction Hamiltonian.\\ 
\subsubsection{Bulk diamond device}
\begin{figure*}[btp!]
  \centering
  \includegraphics[width= 120 mm]{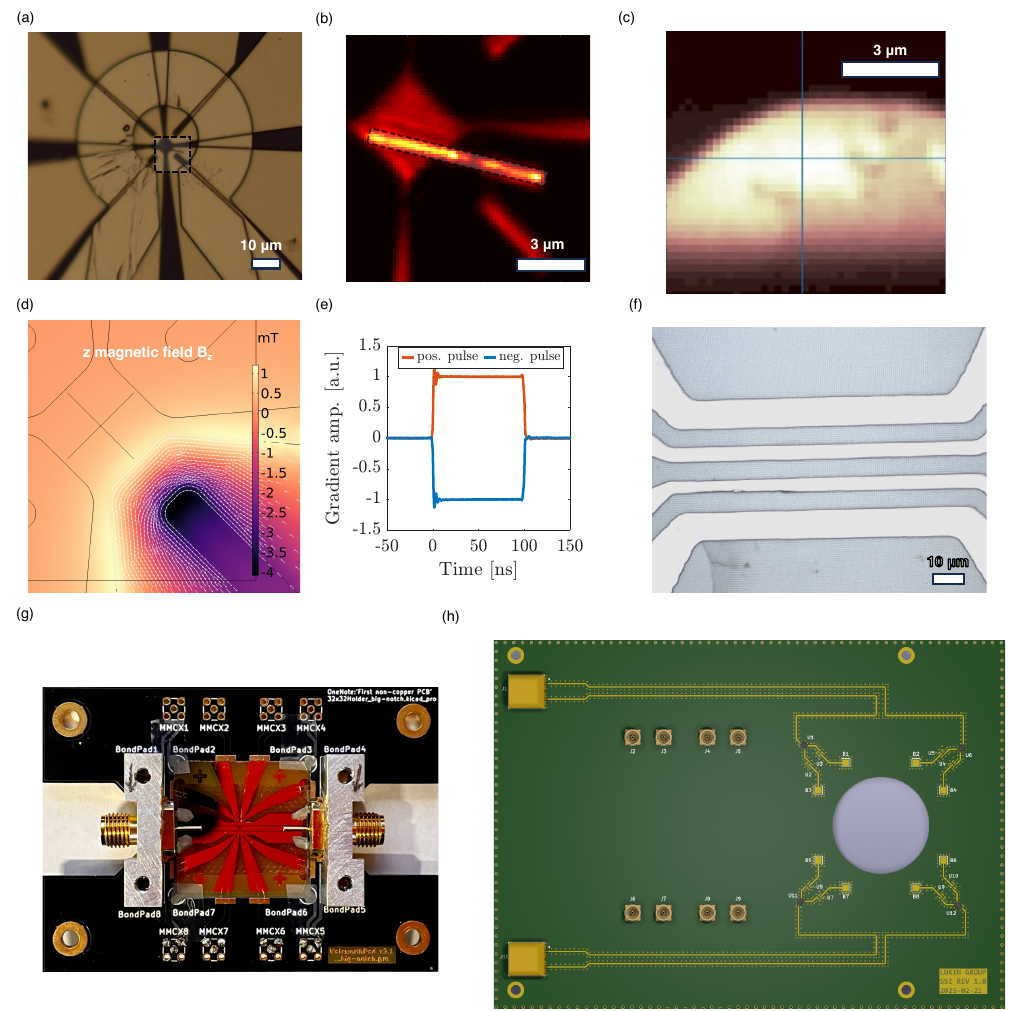}
  \caption{Dynamic magnetic field gradients on the chip.
(a) White-light image of the custom chip providing switchable magnetic gradients and MW control in a diamond nanobeam device. The thin, long beam is a diamond sample hosting
the dipolar ensemble used in parts of this study.
(b) Confocal image of the nanobeam on top of a gradient coil, highlighted with white light. 
(c) Confocal scan of the bulk device showing the dense NV spot between central microcoil wires. 
(d) Calculated gradient field at 1/10 maximum strength in the nanobeam device. 
(e) Dynamic control of the gradient field. Measured current traces for 100-ns positive and negative pulses. 
(f) Microscope image of the four-wire single-layer chip for bulk device; inner pairs yield stronger gradient, outer pairs more linear one. 
(g) Photograph of the two-layer gradient chip on a PCB in the nanobeam device. 
(h) PCB interfacing MW signals and gradient pulses for the single-layer chip, with central hole for laser access.}

\label{fig: SI2}
\end{figure*}

To apply a microwave drive and a switchable gradient to bulk diamond, a single-layer chip was fabricated. The design (Fig.~\ref{fig: SI2}(f)) consists of two pairs of differential lines narrowing down in the central 50~\textmu m region to four gold wires, with widths of 10, 5, 5, and 10~\textmu m, and with equal gaps of 6~\textmu m between the wires. In the experiment, only a single pair of coils is used at a time: the inner pairs are suitable for producing the strongest magnetic field gradient, while the outer pairs yield a more linear gradient. The coil was fabricated in the following steps:
\begin{itemize}
    \item Substrate preparation: A polycrystalline CVD diamond substrate (25 mm x 25 mm x 0.13 mm) was sonicated in acetone first, followed by isopropanol (IPA) sonication.
    
    \item Coil fabrication: Two layers of photoresist (LOR 20B followed by S1813) were spin-coated onto the substrate. The coil pattern was defined by photolithography and developed in MF-319. A metal stack of titanium (Ti, 20 nm), gold (Au, 860 nm), and titanium (Ti, 20 nm) was deposited by electron beam evaporation. Liftoff was performed in 80$^{\circ}$C Remover PG.
    
    \item Dielectric top-layer deposition: An alumina (Al$_2$O$_3$) layer (40 nm) was deposited by atomic layer deposition (ALD) for extra protection against electrical breakdown.

    \item Contact pad opening: A layer of photoresist (S1818) was spin-coated onto the substrate. An etch mask for the bond pads was defined by photolithography and developed in MF-319. The exposed dielectric layer and top titanium layer were removed by buffered oxide etch (1:7 BOE). A final cleaning step was performed in Remover PG, followed by a rinse in acetone and IPA.
\end{itemize}
\subsection{Interfacing PCB}
Combing strong magnetic control and microwave drive requires an interfacing printed circuit board
(PCB).\\
\subsubsection{Nanobeam device}
The two-layer chip is mounted on a simple PCB (see Fig.~\ref{fig: SI2}(g)) that interfaces the gradient and MW lines with the microcoil structure. For MW connections, SMA connector pins are electrically connected to the coplanar waveguide section of the chip using silver paste. Gradient coils are connected via wire bonds from the exposed contact pads on the chip to corresponding pads on the PCB, which are in turn connected to pins interfacing with the gradient pulser. \\
\subsubsection{Bulk diamond device}
For the single layer chip MW control signals are combined with the gradient pulses using a custom-built PCB (see Fig.~\ref{fig: SI2}(h)). The PCB has a symmetric design: the 180°-shifted MW signals (Fairview Microwave FMCP1155 SMA 180 Degree Hybrid Coupler) are fed into two ports on one side of the PCB, while the pair on the other end is terminated. The microwave signal passes through two baluns (TTM Balun Xinger X4BD40L1-50100G) and is then combined using four RF diplexers (Mini-Circuits LDPW-162-242+) with a lower-frequency gradient signal.

\subsection{Diamond sample}
The spin ensemble used in this work consists of spin-1 nitrogen-vacancy (NV) centers in diamond, where a ground-state transition $m_S=-1 \rightarrow m_S=0$ is driven, resulting in an effective spin-1/2 ensemble. Our work uses two different high-NV-density diamond samples.\\
\subsubsection{Bulk sample}
For a most of this work, a bulk, CVD grown, highly doped diamond sample is used (Fig.~\ref{fig: SI2}(c)). The diamond sample was created in a following  process.
Diamond homoepitaxial growth and nitrogen doping were performed via plasma-enhanced chemical vapor deposition (PECVD) using a SEKI SDS6300 reactor on a (100) oriented electronic grade diamond substrate (Element Six Ltd.). Prior to growth, the substrate was fine-polished by Syntek Ltd. to a surface roughness of $\sim$200-300 pm, followed by a 4-5~\textmu m etch to relieve polishing-induced strain. The growth conditions consisted of a 750 W plasma containing 0.5$\%$ $^{12}$CH$_{4}$ in 400 sccm H$_2$ flow held at 25 torr and $\sim$730 $^{\circ}$C according to a pyrometer. A 125 nm-thick isotopically purified (99.998$\%$ $^{12}$C) buffer layer was grown, followed by a 185 nm-thick $^{15}$N-doped layer (1 sccm $^{15}$N$_2$ gas), and a 100 nm-thick $^{12}$C capping layer. After growth, the sample was characterized with secondary ion mass spectrometry (SIMS) to estimate the isotopic purity and epilayer thickness. 
The diamond was further electron irradiated and annealed to generate enhanced NV center concentrations. Irradiation was performed with the 200~keV electrons of a transmission electron microscope (TEM, ThermoFisher Talos F200X G2 TEM). The irradiation time was varied to create spots that range in dose from $10^{17}$-$10^{21}$ e$^{-}$/cm$^{2}$, and the reported experiments are performed at one spot with irradiation dose $2.4\times10^{19}$ e$^{-}$/cm$^{2}$. The sample then underwent subsequent annealing at $850^\circ$C for 6 hours in an Ar/H$_2$ atmosphere, during which the vacancies diffuse and form  NV centers. After irradiation and annealing, the sample was cleaned in a boiling triacid solution (1:1:1 H$_2$SO$_4$:HNO$_3$:HClO$_4$) and annealed in air at $450^\circ$C to oxygen terminate the surface and help stabilize the negative NV$^{-}$ charge state for further measurements.

The density of NV centers in the confocal spot is estimated based on the XY16 decay timescale  of 21.5~\textmu s, corresponding to a single-group NV$^-$ density of 246~ppb. The conversion between XY16 decay timescale and NV$^-$ density is obtained empirically based on numerical simulations, assuming that the decay is dominated by dipolar interaction between NV centers.\\
\subsubsection{Nanobeam sample}
A piece of \textit{black diamond}, ($\sim$3.8~ppm per NV group) characterized in earlier works~\cite{kucsko_critical_2018} is used for the later part of this work (Fig.~\ref{fig: SI2}(b)). The high density of NVs leads to strong magnetic dipole coupling ($J_{\text{typ}} \approx 35$~kHz) between spins, while strain and the presence of other defects result in strong on-site disorder ($W_{\text{typ}} \approx 4$~MHz), necessitating the use of decoupling sequences. The diamond sample is shaped into a triangular nanobeam (0.3~$\times$~8~\textmu m) and placed on top of the chip (see Fig.~\ref{fig: SI2}(a),~(b)). Shaping the diamond into a nanostructure improves fluorescence collection, enhances the homogeneity of the Rabi drive, and enables the application of strong gradients with a well-defined direction $\bm{\nabla}$. 

\subsection{Initialization and readout}
The NV spin state is initialized and read out using a custom-built confocal microscope operating under ambient conditions. Green laser light (532~nm) is focused onto the sample through a high-NA objective. The device is mounted on a piezo stage, which is used to control the position of the confocal spot and allows for $\hat{z}$-focusing. Red NV fluorescence is collected through the same objective and reflected by a dichroic mirror (which filters out the excitation light) towards a single-mode fiber acting as a pinhole to reject out-of-focus fluorescence. The collected fluorescence is then focused onto a pair of single-photon counting modules (bulk sample) or a multi-pixel photon counter module (nanobeam diamond) for measurements.

\subsection{MW control}
\subsubsection{Bulk diamond device}
For the bulk diamond device, MW pulses  ($f = 2.3743$~GHz) are directly synthesized using an arbitrary waveform generator (AWG) Tektronix AWG7122C and amplified using Mini-Circuits ZHL-16W-43-S+. \\
\subsubsection{Nanobeam device}
For a nanobeam device, MW pulses ($f = 2.5036$~GHz) are generated via IQ mixing (Marki Microwave MMIQ-0205H) of signals from a microwave generator (Rohde\&Schwarz SMC100A) and analog control pulses synthesized by the AWG (Tektronix AWG 7052). After IQ mixing, the MW signal is lowpass filtered ($<3.2$~GHz) to eliminate spurious harmonics and then amplified (Mini-Circuits ZHL-25W-63+). IQ leakage is minimized at the qubit frequency to prevent spurious driving between pulses.

\subsection{Gradient control}
\begin{figure}
  \centering
  \includegraphics[width=\linewidth]{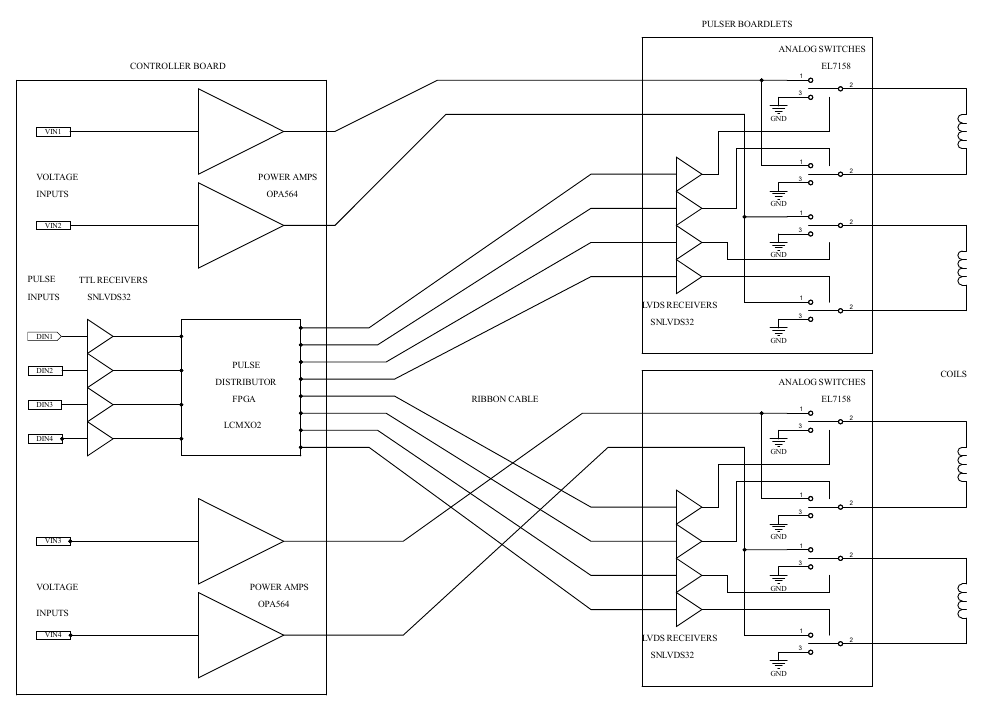}
  \caption{Gradient pulser design. Schematics of the pulser used to create switchable magnetic field gradients.
  }
\label{fig: SI3}
\end{figure}
The switchable gradient field is controlled by a custom FPGA-based current pulser, enabling short ($>5$~ns) current pulses with switchable polarity and an amplitude up to $\sim$ 1.1 A (Fig.~\ref{fig: SI2}(e)). The pulser consists of three main components (Fig.~\ref{fig: SI3}). First, the main board (controller board) generates low-noise, stable voltage references for each coil using ultralow-noise, ultrahigh-PSRR linear regulators (Analog Devices LT3045), which are subsequently amplified by power operational amplifiers (Texas Instruments OPA564). On the same main board, the supplied TTL logic signals are processed using a programmable FPGA (Lattice Semiconductor Corporation LCMXO2), enabling flexible control over the polarity and switching of the pulses. Finally, two small daughter boards are connected to the main board, each containing a pair of non-inverting quad CMOS drivers (Renesas EL7457), which serve as high-current, high-speed analog switches.
The current pulses are triggered using marker channels from a high-sampling-rate AWG , allowing synchronization of gradient and MW pulses. 

\subsection{Addressing selected NV group}
To address the $m_S=-1 \rightarrow m_S=0$ transition in a particular crystallographic group of NVs, an external static magnetic field is applied to split the Zeeman levels of the NV ground state. The magnetic field is generated by a set of three perpendicular electromagnetic coils, and repeated electron spin resonance (ESR) experiments are used to finely align the field direction with the quantization axis of each NV group. This alignment information is also used to transform the NV group orientations into the lab frame, which allows for the reconstruction of the relevant experimental geometry—namely, the orientation of $\bm{\eta}$ for each NV group. To improve readout contrast, the laser light polarization is adjusted individually for each NV group via a half wave-plate.

\subsection{Determination of gradient directions}
\begin{figure}
  \centering
  \includegraphics[width=\linewidth]{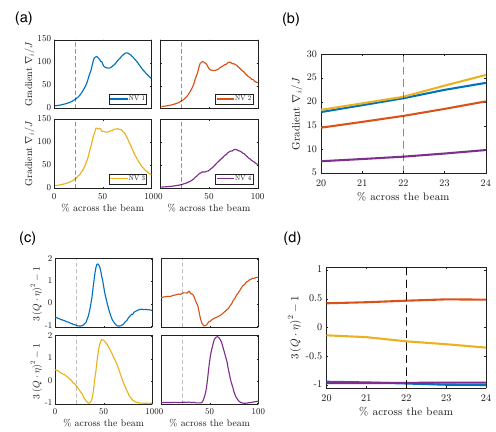}
\caption{Magnetic gradient geometry for a nanobeam device. (a) Gradient strength, in units of typical dipolar spin-spin coupling, across the nanobeam for all four NV groups. (b) Gradient strength in a narrow range near the working point. (c) Geometric factor determining the sign and strength of the dipolar mean field across the nanobeam for all four NV groups. (d) Same geometric factor near the working point. Dashed vertical lines corresponds to to a working point on the nanobeam - $22\% $ along its length.}
\label{fig: SI4}
\end{figure}
\subsubsection{Bulk diamond device}
We determine the geometry of the pulsed magnetic field in the bulk sample device by measuring a series of ESR spectra with electric current applied to individual coils, as a function of position across the middle two wires near the working spot on the sample (see Fig.~\ref{fig: 1}(e)). By fitting the ESR spectra for all NV groups, we reconstruct the vector magnetic field at each spatial position and for each current configuration. The resulting position-dependent vector magnetic field is further fitted to a local quadratic model, to extract the field gradient within the NV plane. The out-of-plane components of the gradient are then inferred based on constraints from Maxwell's equations. This calibration procedure allows us to determine the appropriate current ratios in the individual wires required to achieve a desired magnetic field gradient direction and strength.\\
\subsubsection{Nanobeam device}
To determine the geometry of the magnetic field on a nanobeam device we use finite element method simulations of the magnetic field produced by the microcoil (Fig.~\ref{fig: SI2}(d)). Additionally, we collect a series of confocal scans under extra white light illumination, which highlights the edges of the microcoil (Fig.~\ref{fig: SI2}(b)), to extract the lateral and axial (z) position of the diamond nanobeam relative to the microcoil structure. We then use the simulated magnetic field and the best estimates of the beam position as input to an optimization procedure that compares the experimentally measured ESR detunings along the nanobeam with the predictions from simulations (see Fig.~\ref{fig: 1}(f)). In this optimization, we allow only for a translation of the nanobeam relative to the coil center and an overall scaling of the magnetic field strength. This procedure results in a shift of less than 0.6 {\textmu m} in the nanobeam position compared to the values extracted directly from the confocal scans. In this way, we obtain the magnetic field model used to determine the direction and strength of the effective gradient $\bm{\nabla}_\alpha$ used throughout this work (see Fig.~\ref{fig: SI4}).

\subsection{Preparation of spin spirals}
In this work, we prepare spin spirals by evolving NVs in an inhomogeneous magnetic field created by the gradient coils. In the nanobeam device, the strength of the gradient ($\nabla_{i}$) varies across the beam and NV groups and is plotted in Fig.~\ref{fig: SI4}(a),~(b) in units of dipole interaction strength. The interplay between interaction and gradient strength is important for the preparation of high-quality spin spirals, see Fig.~S3(d),~(f). 

Spiral winding is implemented in two blocks separated by a $\pi$-pulse on the spins, which allows for decoupling of static on-site magnetic disorder that would otherwise dominate over the local gradient field. To avoid transient gradient effects (see Fig.~\ref{fig: SI2}(e)) from impacting the microwave drive, an additional padding time (10~ns on the nanobeam device, and 50~ns on the bulk sample device) per rise/fall edge of the gradient is applied between the gradient pulses and neighboring MW pulses.

\subsection{Hamiltonian engineering}
\begin{figure*}[bpt!]
  \centering
  \includegraphics[width=180 mm]{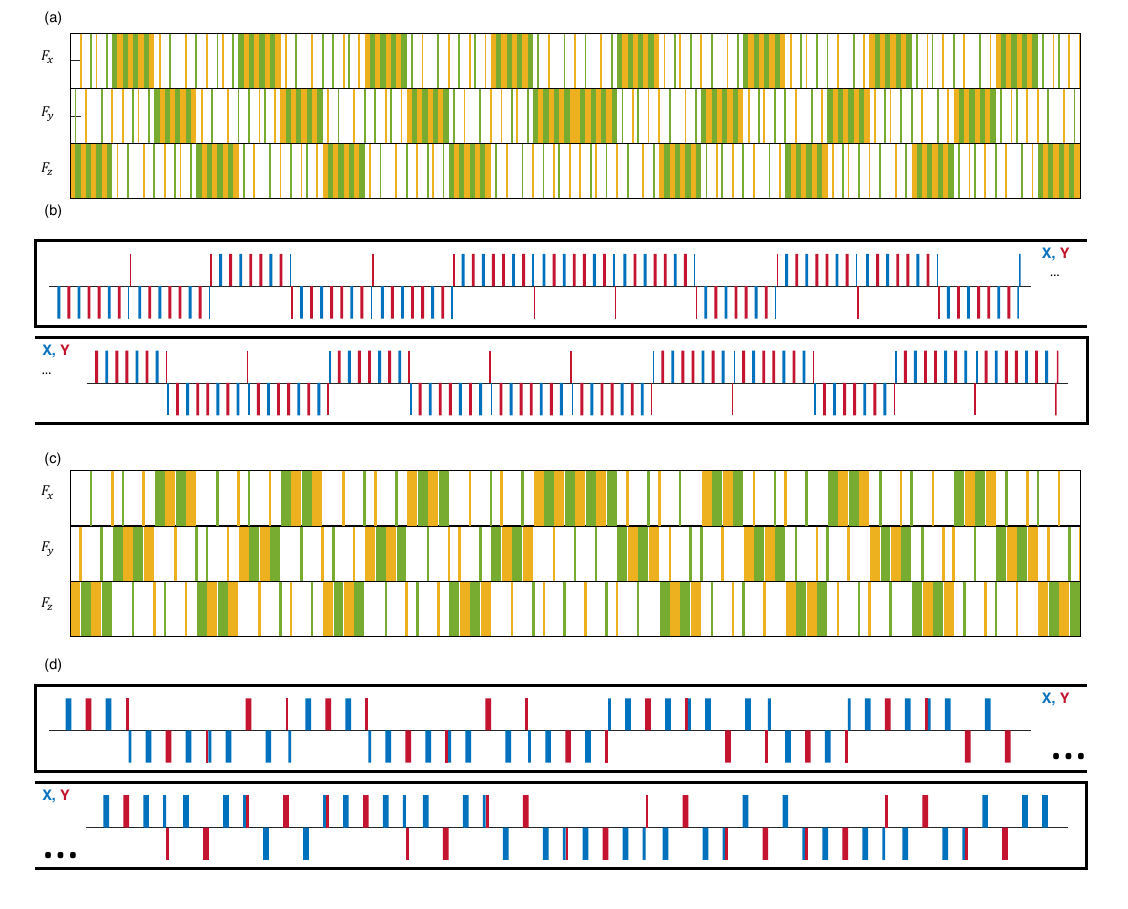}
    \caption{Hamiltonian engineering pulse sequences. (a) Frame representation \cite{choi_robust_2020} of the pulse sequence ``cXY8-DROID-vXY4-Mirror" used for the bulk sample device, showing the concatenated structures discussed in the text.  (b) The actual pulses constituting this sequence. The thin lines represent $\pi/2$-pulses and the thick lines represent $\pi$-pulses. The colors of the pulses represent the pulse axes (X or Y), and the direction of them (up or down) represent the two opposite rotation directions (e.g. $+\pi/2$-pulse and $-\pi/2$-pulse). The ellipsis in the plot indicates that the two rows are connected. The plot is a conceptual illustration of the pulse sequence and is not drawn in proportion to the actual time duration. The actual pulse sequence applied in the experiment uses a cosine envelop with $\pi$-pulse duration $t_\pi=40$~ns and pulse spacing $\tau = 10$~ns. (c,~d) Similar plots for the sequence ``cXY4-DROID-vXY4-Sym" used for the nanobeam device. The actual pulse sequence applied in experiments uses a Gaussian envelope with $\pi$-pulse duration $t_\pi=30$~ns and pulse spacing $\tau = 30$~ns.}
\label{fig: SI5}
\end{figure*}

To engineer a desired form of interaction in a spin ensemble, we make use of Floquet engineering techniques described in detail in \cite{martin_controlling_2023}. The native form of NV-NV interaction (considering $m_S=-1$, $m_S=0$ sublevels),
\begin{equation}\label{eq: native-interaction}
    H_{int}=\sum_{ij} J(\bm r_{ij})\left(S_{i}^{x}S_{j}^{x} +S_{i}^{y}S_{j}^{y}-S_{i}^{z}S_{j}^{z}\right)
\end{equation}
is transformed by a sequence of $\pi/2$ and $\pi$-pulses to a Heisenberg Hamiltonian. By changing pulses spacings, we can continuously tune the form of interaction, as shown in \cite{martin_controlling_2023}. 
In the bulk sample part of this work, we use a new pulse sequence called ``cXY8-DROID-vXY4-Mirror" (see Fig.~\ref{fig: SI5}(a),~(b)), which is an improved version of ``DROID-R2D2" sequence introduced in \cite{tyler_higher-order_2023}, with better robustness against coherent pulse errors.

The name of this pulse sequence stands for its structure, which involves the concatenation \cite{khodjasteh_fault-tolerant_2005} of the following four pulse sequence layers:
    \begin{itemize}
        \item The inner layer is an XY8 sequence, targeting at robust disorder decoupling on fastest possible timescale.
        \item The second layer is the DROID \cite{martin_controlling_2023} structure that tunes the XXZ anisotropy of the effective Hamiltonian.
        \item The third layer can be viewed as further concatenation with XY4 using virtual pulses \cite{alvarez_iterative_2012}. This structure provides improved robustness against coherent pulse errors, as errors accumulated in the first two layers are coherently canceled in this layer. We note that such concatenation with XY4 automatically guarantees the satisfaction of all design rules for higher order dynamical decoupling \cite{tyler_higher-order_2023}.
        \item The outer layer is a mirror symmetrization (in term of the frame representation in Fig.~\ref{fig: SI5}(a)) that we found helpful experimentally. We note that previous DROID-type sequences \cite{choi_robust_2020,tyler_higher-order_2023} also have the same or similar structures.
    \end{itemize}
    
In this work, the sequence described above results in significant extension of experimental timescales under the SU(2) symmetric Heisenberg Hamiltonian, when compared to ``DROID-R2D2" \cite{zhou_robust_2023}, as seen in Fig.~S1(a),~(b).

Similarly, in the nanobeam part of this work, we use another new pulse sequence called ``cXY4-DROID-vXY4-Sym", where the inner layer is replaced by an XY4 sequence to shorten the total sequence duration, and the outer layer is replaced by a slightly different symmetrization. The comparison of experimental timescales to  ``DROID-R2D2" is shown in Fig.~S1(c),~(d).

\subsection{Numerical simulations of spin dynamics}
Numerical simulation of spiral dynamics is done using the discrete truncated Wigner approximation (dTWA) \cite{schachenmayer_many-body_2015}, assuming open boundary condition and experimentally motivated sample geometries. Specifically, we simulate $N=1572$ randomly placed spins for the bulk sample assuming a cylinder geometry of diameter $d=500~$nm and height $h=185~$nm, and $N=6750$ spins for the nanobeam with a length double the beam-width ($w=300~$nm). For all numerics, the UV cut-off spatial scale of NV centers is assumed to be $0.2$ times the typical spacing, (i.e., $r_{\text{UV}}= 6~$nm for the bulk sample and 2.2~nm for the nanobeam).
\section{Acknowledgements}

We are grateful to John Blanchard, Soonwon Choi, Johannes Cremer, Emily J. Davis, Oriana Diessel, Alexander Douglas,  Nazlı Köylüoğlu, Nishad Maskara, Shantam Ravan, Michele Tamagnone,  Yaroslav Tserkovnyak, Zilin Wang, Weijie Wu, Norman Yao, Bingtian Ye, and Qian-Ze Zhu for useful discussions. We thank James MacArthur for technical contributions. We also thank Junichi Isoya, Shinobu Onoda, Hitoshi Sumiya and Fedor Jelezko for providing the \textit{black diamond} sample. This work was supported in part by the Center for Ultracold Atoms (an NSF Physics Frontiers Center), the National Science Foundation (grant number PHY-2012023), Vannevar Bush Faculty Fellowship Program, the Army Research Office through the MURI program grant number W911NF-20-1-0136, Gordon and Betty Moore Foundation Grant No. 7797-01.

\section*{Author contributions}
NTL conceived the mechanism of collective nonlinearity. PP and HG performed the experiment with support from LSM, OM, and HZ. CS, OM, and BM developed the two layer gradient device used to control spin dynamics in the black diamond nanobeam. HG, ACM, PP and MM developed the single layer gradient device used to control spin dynamics in the bulk diamond sample. LBH fabricated the bulk diamond sample. NTL conducted the theoretical analysis with input from HG, PP, LSM and MDL. SP, ACBJ, FC, HP, and MDL supervised all the work. All authors discussed the results and contributed to the manuscript.

\section*{Declarations}
There are no competing interests to declare.

\bibliography{NonlinearDynamics.bib}

\end{document}


\title{Supplementary Materials for \\  ``Collective many-body dynamics in a solid-state quantum sensor controlled through nanoscale magnetic gradients"}
\date{\today}
\affiliation{Department of Physics, Harvard University, Cambridge, Massachusetts 02138, USA}
\affiliation{School of Engineering and Applied Sciences, Harvard University, Cambridge, Massachusetts 02138, USA}
\affiliation{Materials Department, University of California, Santa Barbara, CA 93106, USA}
\affiliation{Harvard Quantum Initiative, Harvard University, Cambridge, MA 02138, USA}
\affiliation{Marian Smoluchowski Institute of Physics, Jagiellonian University in Krak\'ow, 30-348 Krak\'ow, Poland}
\affiliation{Department of Physics, University of California, Santa Barbara, CA 93106, USA}
\affiliation{Department of Chemistry and Chemical Biology, Harvard University, Cambridge, Massachusetts 02138, USA}

\author{Piotr Put$^1$}
\thanks{These authors contributed equally to this work}
\author{Nathaniel T. Leitao$^1$}
\thanks{These authors contributed equally to this work}
\author{Haoyang Gao$^{1}$}
\thanks{These authors contributed equally to this work}
\author{Christina Spaegele$^2$}
\thanks{These authors contributed equally to this work}
\author{Oksana Makarova$^{1,2}$}
\author{Lillian B. Hughes Wyatt$^{3}$}
\author{Andrew C. Maccabe$^{4}$}
\author{Matthew Mammen$^{4}$}
\author{Bartholomeus Machielse$^{1}$}
\author{Hengyun Zhou$^{1}$}
\author{Szymon Pustelny$^{1,5}$}
\author{Ania C. Bleszynski Jayich$^{6}$}
\author{Federico Capasso$^{2}$}
\author{Leigh S. Martin$^1$}
\author{Hongkun Park$^{1,7}$}
\author{Mikhail D. Lukin$^{1}$}

{
\let\clearpage\relax
\maketitle
}
\tableofcontents


\section{Supplementary Material}

\begin{figure}
  \centering
  \includegraphics[width=\linewidth]{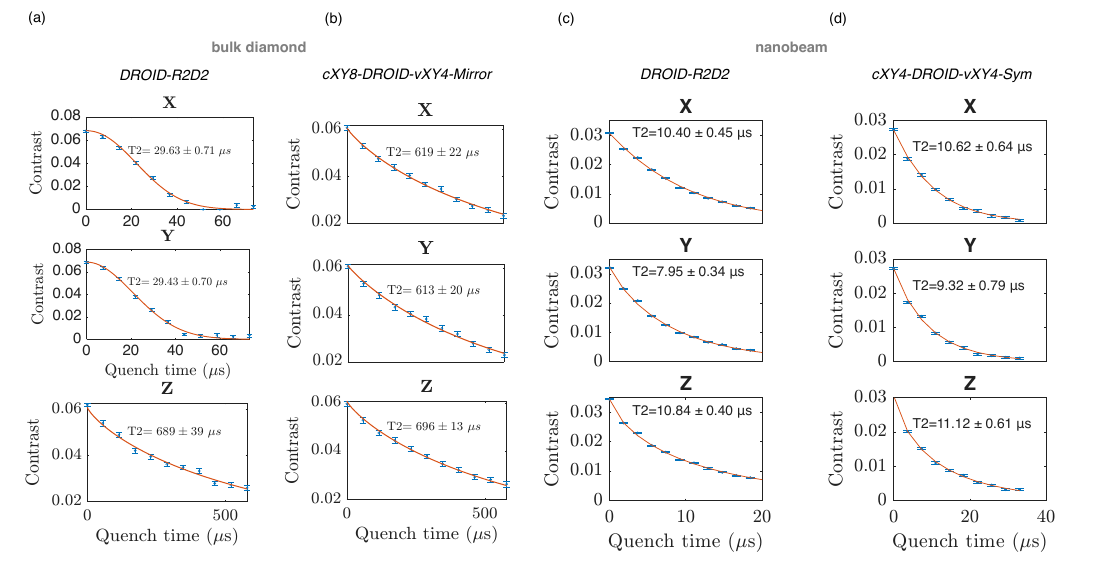}
  \caption{Experimental coherence times. (a) Measured coherence decay of a bulk diamond for initial $\hat{x}$, $\hat{y}$, and $\hat{z}$ states under the DROID-R2D2 decoupling sequence (introduced in \cite{zhou_robust_2023}). Decoherence of the $\hat{x}$ and $\hat{y}$,states is dominated by microwave inhomogeneity. (b) Same measurement with a new symmetrized decoupling sequence (cXY8-DROID-vXY4-Mirror), which engineers the Heisenberg Hamiltonian in the bulk sample. (c) Measured coherence decay of NV group 1 for initial $\hat{x}$, $\hat{y}$, and $\hat{z}$ states under the DROID-R2D2 decoupling sequence in the nanobeam sample. (d) Same measurement with a new symmetrized decoupling sequence (cXY4-DROID-vXY4-Sym), which engineers the Heisenberg Hamiltonian in the nanobeam sample.}
\label{fig: SI6}
\end{figure}

\subsection{Nanophotonics model}
\begin{figure}
  \centering
  \includegraphics[width=0.85\columnwidth]{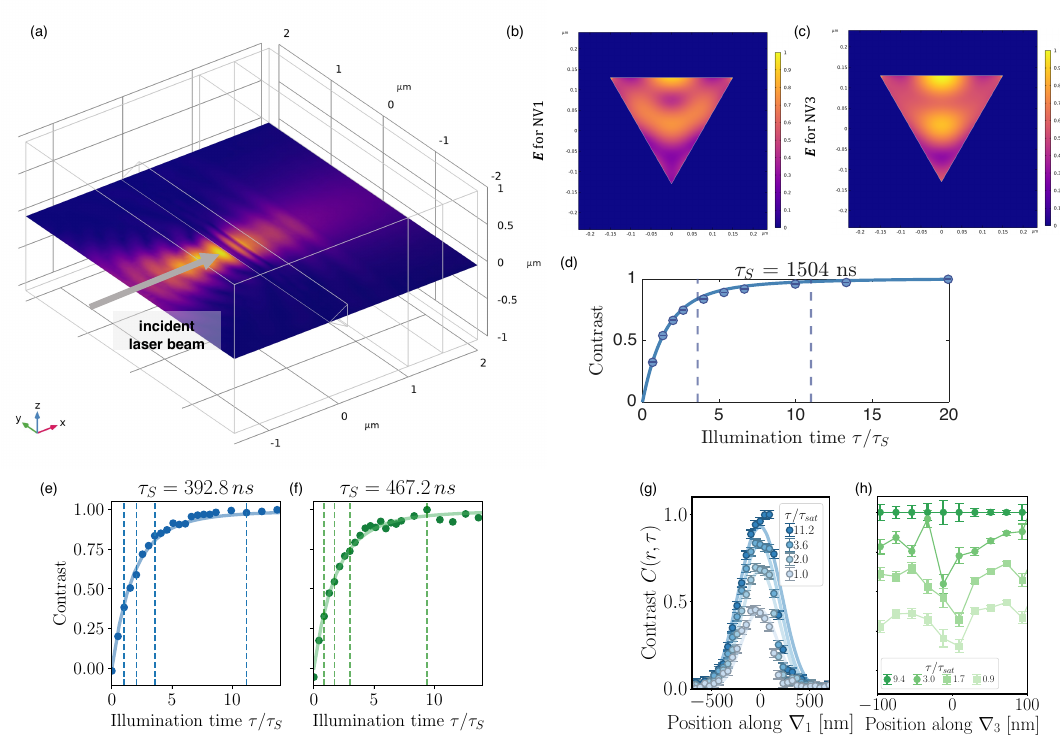}
  \caption{Nanophotonics model of the diamond nanobeam and effects of optical pumping on NV polarization. (a) Simulated model of the diamond nanobeam hosting NVs, placed on a chip and illuminated by a green laser beam. The electric field amplitude $|E|$ is plotted in a plane containing the long axis of the nanobeam. (b,~c) Cross-sectional view showing the electric field amplitude $|E|$ inside the diamond nanobeam, revealing a clear interference pattern of 532~nm laser light used to polarize and readout spins. (d) Measured optical contrast for bulk diamond as a function of total green illumination time. Solid lines show fits to the saturation model. The extracted saturation value $\tau_s$ is used to calibrate measurements in Fig.~4 of the main text. Vertical dashed lines correspond to the optical pumping values used in Fig.~4(c). (e) Same measurement for a nanobeam device using NV1 and (f) NV3 groups. Light polarization for each NV group was optimized to achieve maximum optical contrast. (g) FMIs for NV group 1 along the nanobeam sample; markers are measurements and lines are theoretical predictions from the nanophotonics model. (h) Relative FMIs across the nanobeam (experimental data from the main text, Fig.~4(f)) for different polarization times for NV group 3 in the nanobeam. Since the excitation and collection profiles affecting the FMIs are effectively divided out, the resulting contrast serves as a direct proxy for NV polarization. Shorter optical pumping times lead to the emergence of a ``hole'' in the NV polarization distribution.}

\label{fig: SI7}
\end{figure}

To obtain a realistic model for NV polarization inside the diamond nanobeam, we performed a finite element method (COMSOL Multiphysics) simulation of light propagation through the diamond nanostructure (Fig.~\ref{fig: SI7}). A Gaussian beam ($\lambda_{0} = 532$~nm), focused at its center, propagates perpendicular to the long axis of the nanobeam. The resulting time-averaged electric field distribution is shown in Fig.~\ref{fig: SI7}(b),~(c). A clear interference pattern is observed inside the nanobeam. This effect is responsible for the polarization ``hole" seen in the Fourier Magnetic Image (FMI) taken across the nanobeam (see main text, Fig.~4(f)).
\subsection{Optical pumping rate}

To study the microscopic distribution of spin polarization in diamond nanobeam we take FMIs at different optical pumping rates (see Fig.~4(f) in the main text). The saturation time was extracted from the experimental data measuring NV contrast $C(\tau)$ as a time of green illumination and fitting the nanophotonics model combined with a simple saturation model,
\begin{align}  \label{eq: polarization-model}
    C(\tau) &= \int d\bm r \, w(\bm r) \, \rho(\bm r, \tau), \\
    \rho(\bm r, \tau/\tau_S) &= \rho_0 \left(1-\exp{-\tau |E(\bm r)|^2/\tau_S} \right),
\end{align}
    to the experimental data (see Fig.~\ref{fig: SI7}(e),~(f)). Here, $\rho_0$ is the maximum NV polarization density, and $|E(\bm r)|^2$ is the electric field intensity of green light, and $w(\bm r)$ is a weighting function given by the product of green illumination and red fluorescence intensity obtained from the COMSOL simulations. Data were taken using 35 \textmu W of green laser light focused onto a confocal spot. Low laser power was chosen to minimize light-induced charge effects present in dense ensembles of NV centers \cite{choi_depolarization_2017}. We similarly calibrated the saturation time for the bulk diamond sample (Fig.~\ref{fig: SI7}(d)) using a simple optics model for green light distribution (at 90~\textmu W), assuming a Gaussian beam profile inside the diamond:
\begin{align}
    |E(\bm r)|^2 &= a e^{-\kappa_0^2(z)\rho^2}; \qquad \kappa_0^2(z) = \left(w_0^2+\left(\frac{\lambda_0}{\pi w_0} \right)^2 z^2 \right)^{-1},
\end{align}
    where $\lambda_0$ is the wavelength of green light. 

Due to the cylindrical symmetry, the saturation curve can be calculated analytically (assuming a uniform aspect ratio of intensity ratio across the thickness of the beam)
\begin{align}
    C(\tau) &= a^2\int_{-L_z/2}^{L_z/2} dz \int_0^\infty \rho \, d\rho \, e^{-\kappa^2_0(z)\rho^2} \, e^{-\kappa^2_1(z) \rho^2} \left(1-\exp{\left(-\tau a e^{-\kappa^2_1(z)\rho^2}/\tau_S\right)} \right) \notag\\
    &= \int_{-L_z/2}^{L_z/2} dz \, \int_0^1 du u^{\kappa^2_0(z)/\kappa^2_1(z)}\left(1-e^{-\tau u/\tau_S } \right) \notag\\
    &= \int_{-L_z/2}^{L_z/2} dz \left(1-\frac{\Gamma(\kappa_0^2(z)/\kappa_1^2(z))}{\tau^{1+\kappa_0^2(z)/\kappa_1^2(z)}}+ \left( 1+\kappa_0^2(z)/\kappa_1^2(z) \right) E_{\kappa_0^2(z)/\kappa_1^2(z)}(\tau) \right)
\label{eq: fitgaussint}
\end{align}
where $E_n(z)=\int_1^\infty dt \frac{e^{-zt}}{t^n}$ is the exponential integral function, and $\kappa_1(z)$ is associated to the red collection wavelength.  Eq.~\ref{eq: fitgaussint} was then fit to experimental data in Fig.~\ref{fig: SI7}(d) to obtain the optical pumping saturation time.

\subsection{Additional FMIs}
We complement FMIs taken across the nanobeam (see Fig.~4(f)) with one taken approximately along the beam (using NV group 1), clearly observing a decrease in the extent of the polarized region with shorter optical pumping times (Fig.~\ref{fig: SI7}(g)).  
To study the polarization density across nanobeam diamond from experimentally extracted FMIs, we normalize each FMI (main text, Fig.~4(f)) by the one acquired at the longest optical pumping time, as shown in Fig.~\ref{fig: SI7}(h), to ideally cancel the contribution from the optical weighting function $w(\bm r)$. In this transverse polarization cut (along $\bm \nabla_{3}$), a ``hole'' emerges in the polarization profile for shorter pumping durations, consistent with our polarization model derived from nanophotonic simulations (see Fig.~4(f),~(g), main text).

\subsection{Decoherence during spiral winding}
\begin{figure}[h!]
  \centering
  \includegraphics[width=\linewidth]{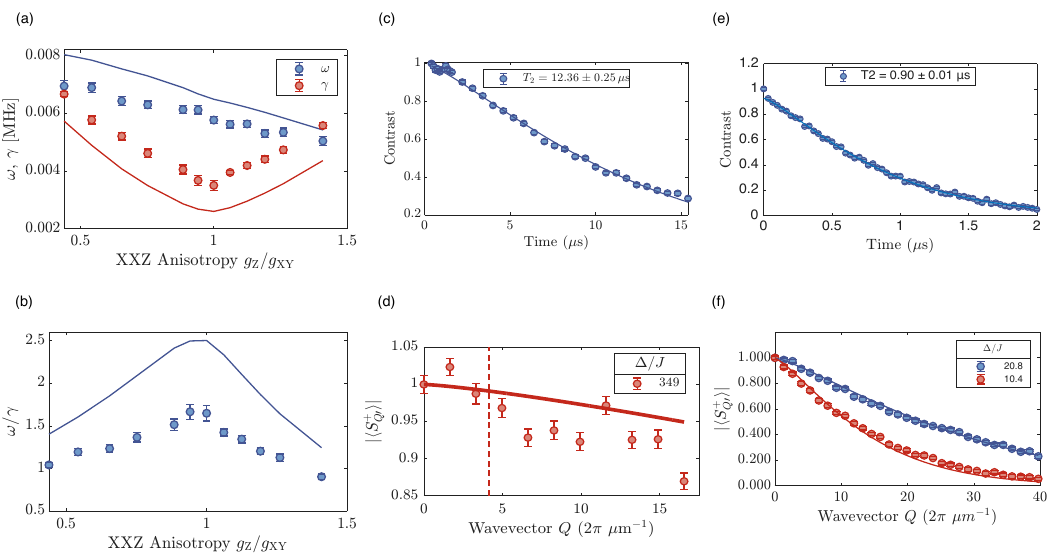}
     \caption{Precession rate and decay of the spiral as a function of XXZ anisotropy and the polarization loss during winding (a) Spiral precession rate $\omega$ and decay $\gamma$ as functions of interaction anisotropy. Markers denote experimental results, while solid lines represent theoretical predictions. The dataset used in this panel is the same as in Fig.~3(g) in the main text. 
    (b) Quality factor as a function of XXZ anisotropy. Markers denote experimental results, while solid line represent theoretical prediction. The worse quality factor than main text Fig.~5(h) is due to the worse gradient linearity and Rabi homogeneity in this measurement.  (c) Polarization decay under a Hahn spin echo (two $\pi$-pulses) for NV spins in a the bulk sample.
    (d) Polarization loss during winding and unwinding for a bulk sample device. Solid lines represent the polarization decay predicted by finite $T_2$ decay, as shown in (c), while the vertical dashed lines indicate typical wavevectors $Q$ used in this work to study spiral dynamics. The ratio of the gradient energy separation between a typical coupled spin pair $\Delta$ and their interaction energy $J$ is indicated.
    (e) Polarization decay under a Hahn spin echo for NV spins in a black diamond nanobeam.
    (f) Polarization loss during winding and unwinding for two different gradient field strengths (the maximum gradient strength and half of it) in a nanobeam device. Solid lines represent the polarization decay predicted by finite $T_2$ decay, as shown in (e).}

\label{fig: SI8}
\end{figure}
In this work, we use a simple Hahn echo sequence with a single $\pi$-pulse to (un)wind the spin spirals. We expect the spin polarization after the (un)winding periods to be limited by the finite $T_2$ time under the spin echo sequence (measured for the bulk sample in Fig.~\ref{fig: SI8}(c) and for the black diamond sample in Fig.~\ref{fig: SI8}(e)). Indeed, when we measure the polarization after immediate winding and unwinding as a function of the spiral wavevector $Q$ (data points in Fig.~\ref{fig: SI8}(d),~(f)), we observe that the loss in polarization follows the $T_2$ decay (overlaid solid lines in Fig.~\ref{fig: SI8}(d),~(f)). For the bulk diamond device, the polarization decay during winding and unwinding of a typical wavevector used in this study is negligible (vertical dashed line in Fig.~\ref{fig: SI8}(d)). For the denser nanobeam sample, this loss becomes significant and is used to accurately predict the additional decay of the spin precession amplitude as a function of $Q$, as illustrated by the difference between the solid and dashed lines in Fig.~4(e) of the main text.
It is worth noting that in the case of lower NV densities or stronger field gradients, technical factors—such as shot-to-shot fluctuations in the gradient pulse current—can cause additional contrast loss during the winding/unwinding periods, thereby limiting spin polarization in many-body physics studies and reducing the achievable resolution in FMIs.

\subsection{Spiral precession quality factor for different interaction anisotropies}
In the main text, Fig.~3(g), we demonstrate that the precession amplitude of the spiral is maximized at the SU(2)-symmetric point of interactions. The measured dependence results from the interplay between the twisting rate and spiral decay, as shown in Fig.~\ref{fig: SI8}(a): the precession rate $\omega$ decreases as the exchange term in the Hamiltonian becomes smaller, while the relaxation rate is minimized at the SU(2) point. This leads to the precession amplitude (Fig.~3(g), main text) and the quality factor plotted in Fig.~\ref{fig: SI8}(b) being peaked at the Heisenberg point, while exhibiting a pronounced asymmetry between the easy-plane and easy-axis sides. Moreover, as seen in Fig.~\ref{fig: SI8}(a), the experimental precession rate is slower than the theoretical prediction by a constant factor, most likely due to errors in the estimated spin polarization and NV density, or an inhomogeneous gradient field. In addition, the experimentally observed decay contains an extra contribution, most likely arising from the finite coherence time (see Fig.~\ref{fig: SI6}(b)) and inhomogeneous microwave drive. These features are likely responsible for the deviations between experimentally measured and theoretically predicted precession amplitudes at larger detunings from the SU(2) point, as shown in Fig.~3(g) in the main text.

\subsection{Theory of spin spiral exchange in dipolar systems}
\begin{figure}
  \centering
  \includegraphics[width=0.55\linewidth]{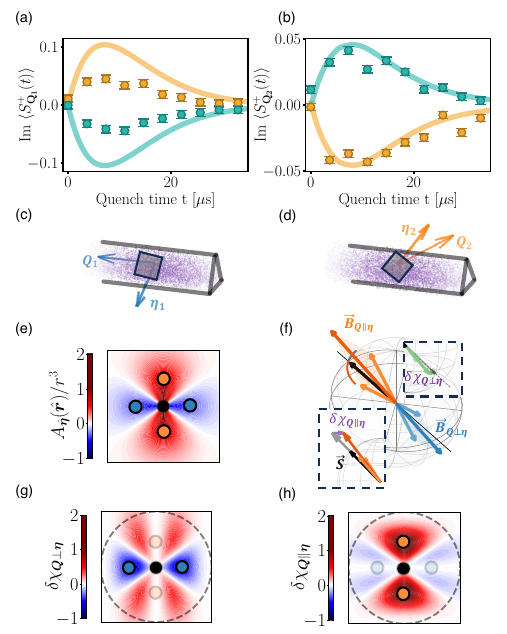}
  \caption{Spiral precession in a nanobeam sample and mean-field mechanism. (a,b) Measured spiral dynamics in NV groups~1 and~2 at $Q = 2\pi \times 8~\text{\textmu m}^{-1}$; markers are data, lines are theory including reduced polarization and decoherence.  (c,d) Experimental geometries in the nanobeam, approximately:  $\bm{\eta}_1 \perp \bm{Q}_1$ and $\bm{\eta}_2 \parallel \bm{Q}_2$.  (e) Illustration of the dipolar interaction sourced by a central spin at the origin (black dot). Two pairs of spins coupled to the central spin are shown, one pair aligned with the dipolar quantization axis (orange), and the second perpendicular (blue). (f) Illustration of the mean-fields sourced by the pair of spins that dominate the spin exchange discussed above, in the frame of the central spin. Insets show the transverse component of the mean field that directly determines the precession frequency. (g) In the case where the spiral wavevector is aligned parallel to the dipolar quantization axis, the pair of spins aligned with this axis sources the dominant component of the exchange mean field. The pair of spins in the plane perpendicular to the pitch of the spiral (and perpendicular to the dipolar quantization axis) does not contribute to precession since their polarization is collinear with the central spin. (h) For spiral $\bm{Q} \perp \eta $, spins aligned with the pitch of the spiral are coupled with the opposite sign of the dipolar anisotropy, and dominate the exchange field.}

\label{fig: SI9}
\end{figure}

In what follows, we consider XXZ Hamiltonians parameterized as
    \begin{align}
        H_\lambda &= \sum_{i<j} J(\bm r_{ij}) \left(g_0(\lambda) \bm S_i \cdot \bm S_j + g_2(\lambda) S_i^z S_j^z \right),
    \end{align}
keeping in mind that the experiment implements $g_0(\lambda) = 2\left( 1+\lambda\right)/3, g_2(\lambda) = -2\lambda$, for $-1< \lambda\leq 2$, at the level of average Hamiltonian theory \cite{martin_controlling_2023}, where $\lambda=0$ is the engineered SU(2)-symmetric point (Eq. (2) in the main text) and $\lambda=2$ is the native interaction, Eq.~4 in the main text. 
Here, the intrinsic dipole-dipole couplings $J(\bm r)$ are set by the dipolar quantization axis $\bm \eta$ of the relevant NV group, 
\begin{align}
    J(\bm r) 
    &\equiv \frac{J_0}{r^3} A_{\hat{\bm \eta}}(\hat{\bm r}) \notag\\
    &= \frac{\mu_0\gamma_{NV}^2\hbar^2}{4\pi r^3} \, \frac{3(\hat{\bm \eta}\cdot \hat{\bm r})^2-1}{2},
\end{align}
where $\mu_0$ is the magnetic permeability, and $\gamma_{NV}$ is the gyromagnetic ratio of a single NV center. Connecting to the conventions in the main-text, we have 
\begin{align}
    \mathcal{A}_{\hat{\bm \eta}}(\hat{\bm r}) &= \frac{2}{3}J_0A_{\hat{\bm \eta}}(\hat{\bm r}),
\end{align}
as a rescaled form of the dipolar anisotropy originating from the invariance of the Hamiltonian norm (Hamiltonian trace) under Floquet engineering. Given this model of the spin interactions, we proceed to analyze the theoretical expectations for many-body dynamics of the spin spiral.
\\
\\
\noindent
\subsubsection{Analytical derivation of exchange field from a mean-field argument}
To build a minimal model for the dynamics observed within the coherence time of our experiment,  we employ the early-time expansion
\begin{align}
    \mathcal{S}_{\bm Q,\theta}(t) &= \langle S^+_{\bm Q}(t) \rangle_{\rho_{\bm Q,\theta}}
     = \sin\theta \, e^{i\Omega_{\bm Q,\theta} t + \mathcal{O}((\mathcal{J}t)^2)},
\end{align}
where spin spiral precession frequency is obtained via 
\begin{align}
    \Omega_{\bm Q,\theta} &=\Im \frac{1}{\langle S_{\bm Q}^+\rangle_{\rho_{\bm Q,\theta}}} \partial_{t} 
    \left\langle S_{\bm Q}^+(t)  \right\rangle_{\rho_{\bm Q,\theta}} \bigg \vert_{t=0} \notag \\
    &=\frac{1}{\langle S_{\bm Q}^+\rangle_{\rho_{\bm Q,\theta}}} \langle \lbrack S^+_{\bm Q},H \rbrack  \rangle_{\rho_{\bm Q, \theta}}. 
\end{align}
Physically, this corresponds to the average torque applied to each spin in the ensemble, normalized by transverse spin polarization,  in the quantum state
\begin{align}
    \rho_{\bm Q,\theta} &= V_{\bm Q} \prod_j \frac{1+P(\bm r_j) \left( \sin{\theta} X_j + \cos{\theta}Z_j\right)}{2} V_{\bm Q}^\dagger,
\end{align}
where $P(\bm r_j)$ is the polarization at site $\bm  r_j$ and 
\begin{align}
    V_{\bm Q} &= \exp{ -i \bm Q \cdot \sum_j \bm r_j \, S_j^z }
\end{align}
is the idealized local gradient rotation unitary. Equivalently, we can analyze this torque in the frame of the spiral, in which all spins are counter-rotated to be polarized in the $XZ$ plane at angle $\theta$. In this rotated frame, the effective Hamiltonian becomes
\begin{align}
    H(g_0,g_2) \to H_{\bm Q}(g_0,g_2) &= V_{\bm Q}^\dagger H(g_0, g_2) V_{\bm Q}\\
    &= \frac{g_0}{2}\sum_{ij} J(\bm r_{ij}) \left( e^{i\bm Q \cdot \bm r_{ij}}S_i^+ S_j^- + e^{-i\bm Q \cdot \bm r_{ij}}S_i^- S_j^+ \right) \notag\\
    & \qquad + \frac{\left(g_0+g_2\right)}{2}\sum_{ij} J(\bm r_{ij}) S_i^z S_j^z,
\end{align}
and the early-time frequency is 
\begin{align}
    \Omega_{\bm Q,\theta} &= \frac{ \langle \lbrack  S_{\bm 0}^+, H_{\bm Q} \rbrack \rangle_{\rho_{\bm 0,\theta}}  }{\langle S_{\bm 0}^+\rangle_{\rho_{\bm 0,\theta}} }.
\end{align}
The dynamics of the spin observables then take a simple form 
    \begin{align} 
     i\partial_t \, S_j^+ &= -\lbrack H_{\bm Q}, S_j^+\rbrack = B_j^z S^+_j - B_j^+(\bm Q) S_j^z, \\
    i\partial_t S_j^z &= -\lbrack H_{\bm Q}, S_j^z\rbrack =B_j^+(\bm Q) S_j^- - B_j^-(\bm Q) S_j^+,
    \end{align}
where we have defined the local mean-field $\bm{B}_j$ with transverse components, $B_j^\pm = B_j^x\pm i B_j^y$
    \begin{align}
        B_j^{\pm}(\bm Q) &= g_0 \sum_l J(\bm r_{jl}) e^{\mp i \bm Q \cdot \bm r_{jl}} S_l^\pm,
    \end{align}
illustrated in Fig.~\ref{fig: SI9}(f), and longitudinal component 
\begin{align}
    B_j^z &= (g_2+g_0)\sum_l J(\bm r_{jl})  S_l^z.
\end{align}
Evaluating the expectation value in the initial state, we obtain the final result
\begin{align}
    \Aboxed{\Omega_{\bm Q}(\theta) = \cos\theta \left( g_0  \, \chi^{XY}_{\bm Q} + g_2 \, \chi^{ZZ} \right).}
\end{align}
Here, we explicitly separate the exchange field strength $\chi_{\bm Q}=\chi^{XY}_{\bm Q}$ that is determined by the integral over the spiral texture, weighted by the dipolar couplings
\begin{align}\label{eq: exchange-field si}
    \chi^{XY}_{\bm Q} &= \frac{1}{\rho_0}\int d\bm r \,  \rho(\bm r) \, \int d\bm r' \rho(\bm r') J(\bm r - \bm r')  \left(1- \cos{\left( \bm Q \cdot (\bm r- \bm r') \right)}\right)
\end{align}
from the longitudinal Ising component
\begin{align}
    \chi^{ZZ} &= \frac{1}{\rho_0}\int d\bm r \rho(\bm r) \int d\bm r' \rho(\bm r') J(\bm r-\bm r'), 
\end{align}
which averages to zero in a three-dimensional dipolar ensemble. Furthermore, note that we have explicitly defined the polarization density as 
\begin{align}
    \rho(\bm r) &= \sum_j P(\bm r) \delta(\bm r-\bm r_j),
\end{align}
and the total spin polarization 
\begin{align}
    \rho_0 &= \int d\bm r \rho(\bm r)
\end{align}
to normalize the exchange fields.

To gain further intuition for this result, we visualize the effect of the dipolar anisotropy in Fig.~\ref{fig: SI9}(e),~(f),~(g),~(h). In particular, we decompose the exchange field  (Eq.~\ref{eq: exchange-field si}), as an integral over pairs of spins in the spiral texture, separated from the central spin with equal and opposite displacements (Fig.~\ref{fig: SI9}(g),~(h)). Due to the equal and opposite displacement, the individual coherences will be rotated by $\pm \bm Q \cdot \bm r$ around the $XY$ plane, producing a total mean-field that is canted slightly above the central spin polarization, as illustrated by the orange and blue arrows in Fig.~\ref{fig: SI9}(f). Crucially, the magnitude of this deflection encodes both the \textit{sign} of the dipolar anisotropy and the \textit{orientation} of the spin pair relative to the spiral wavevector. Spins separated perpendicular to the wavevector of the spiral do not produce any deflection because their polarization is parallel to the central spin. In contrast, spins separated along the spiral wavevector source a larger deflection. 
\\
\\
\subsubsection{Numerical analysis of spin exchange in the overdamped limit}
The dominant limitations to achieving a large spiral precession signal in the nanobeam sample (see Fig.~\ref{fig: SI9}(a),~(b)) arise primarily from the finite coherence times measured in Fig.~\ref{fig: SI6}(d). In addition to this extrinsic dephasing, we investigate the effects of geometry and finite spin polarization using dTWA numerical simulations. In particular, we observe a clear dependence on the relative orientation between the dipolar quantization axis and the spiral pitch (see Fig.~\ref{fig: SI10}(a)), consistent with the experimental results (see Fig.~3(f) in the main text) and the mean-field analysis discussed above (see also Eq.~\ref{eq: isotropic exchange-field}). Furthermore, we estimate the maximum spin polarization for the nanobeam sample to be $P \sim 0.5$, based on measurements of the spin contrast (comparing $3.8\%$ contrast in the nanobeam sample to $6.3\%$ contrast in a dilute NV ensemble measured in the same confocal system, where the spin polarization can theoretically reach $P_{\mathrm{dilute}} \sim 0.85$). Numerical simulations confirm that this significantly reduces the precession amplitude (see Fig.~\ref{fig: SI10}(b). For the bulk diamond sample, we assume a higher polarization $P = 0.8$. 
\\
\\
\subsubsection{Dipolar exchange fields as a probe of microscopic polarization extent}
\begin{figure}
  \centering
  \includegraphics[width=\linewidth]{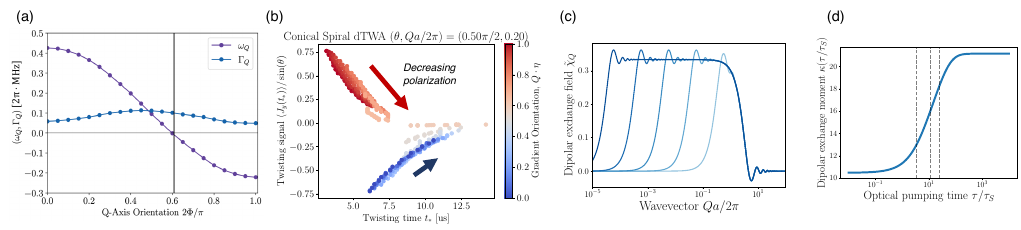}
  \caption{Numerical simulation of exchange dynamics and sensitivity of to microscopic details in a nanobeam device. (a) Effect of geometric orientation on spin precession rate, as calculated by dTWA. Vertical line indicates the magic angle where no spiral precession is expected.  (b) Spiral precession amplitude and the time at which maximum precession occurs, linearly varying the polarization of the spin spiral from $0.5$ to $1.0$. (c) Exchange field sourced to a central spin as a function of $Q$, where different curves index different IR cutoffs, $R_*=10,10^2,10^3,10^4,10^5$. (d)~Exchange moment of the polarization model associated with the nanophotonic simulation of the nanobeam device, which is highly sensitive to the optical pumping times probed in the experiment (vertical dashed lines).}

\label{fig: SI10}
\end{figure}

We now turn to an analysis of the exchange field integral, Eq. \ref{eq: exchange-field si}, in the three-dimensional geometry. For an isotropic polarization distribution, it is straightforward to calculate the exchange field sourced at a central spin as     
\begin{align}\label{eq: isotropic exchange-field}
    \tilde{\chi}_{\bm Q} &=  n \int d\bm r \, \frac{A(\hat{\bm r} \cdot  \hat{\bm \eta} )}{r^3} \left( 1- \cos{\left( \bm Q \cdot \bm r \right)}\right) \notag \\
    &= \frac{4 \pi}{3} A(\hat{\bm Q} \cdot \hat{\bm \eta}) \, n \int_a^{L} \frac{dr}{r}  G(Q r) \notag \\
    &= \frac{4 \pi}{3} A_{\hat{\bm \eta}}(\hat{\bm Q}) \left \lbrack \frac{3\left(\text{sinc}\,\left(Qa \right)-\cos{\left(Qa\right)}\right)}{\left(Qa\right)^2} - \frac{3\left(\text{sinc}\,\left(QR_* \right)-\cos{\left(QR_*\right)}\right)}{\left(QR_*\right)^2}\right \rbrack,
\end{align}
where $n$ is the density of NV centers, and
\begin{align}
    G(x) &= \frac{(3-x^2)\sin{x}/x - 3\cos{x}}{x^2},
\end{align}
is the Fourier transform of the dipolar anisotropy. The resulting function is plotted on a semi-log plot in Fig.~\ref{fig: SI10}(c), where the upper limit of the integral is varied across five orders of magnitude, showing striking sensitivity to this IR-cutoff.

For more general polarization distributions, the precession rate is determined by a convolution with the UV-regulated dipolar interaction $J_{\bm q}$, with an exchange kernel $\mathcal{K}_{\bm Q}(\bm q)$ associated with the polarization distribution. Namely,
\begin{align}
    \chi_{\bm Q} &=  \frac{1}{\rho_0}\int d\bm r \, w(\bm r) \rho(\bm r) \int d\bm r' \rho(\bm r')J(\bm r- \bm r')\left(1-\cos{\left(\bm Q \cdot \left( \bm r - \bm r'\right) \right)}  \right) \notag\\
    &= \frac{1}{\rho_0}\int d\bm q \, \mathcal{K}_{\bm Q}(\bm q) J_{\bm q}; \\
    \mathcal{K}_{\bm Q}(\bm q) &= \left(w \star \rho \right)_{\bm q}^* \rho_{\bm q} - \frac{1}{2}\left((w \star \rho)_{\bm q+\bm Q}^* \rho_{\bm q+\bm Q} + (w \star \rho)_{\bm q-\bm Q}^* \rho_{\bm q- \bm Q} \right).
\end{align}
where $w$ is the optical weighting function and $\star$ denotes convolution in momentum space.

Here, we assume the dipolar interactions are decoupled in the UV with some cutoff $r_{UV}=2\pi\Lambda^{-1}$ associated with a minimal seperation between NV centers, while  $\mathcal{K}_{\bm Q}$ incorporates the IR details of the polarization distribution.  For the dipolar interactions, it is straightforward to verify
\begin{align}
    J_{\bm q} &= -\frac{4\pi }{3}  \, A_{\hat{\bm \eta}}(\hat{\bm q}) \, \mathcal{V}(q/\Lambda),
\end{align}
where $\mathcal{V}(x) = 3\left(\text{sinc}\, x-\cos{x}\right)/x^2$. Crucially in $\bm q$-space, the dipolar interaction is \textit{gapped}
\begin{align}
    \lim_{x \to 0}\mathcal{V}(x) &= 1,
\end{align}
by virtue of the matching of the exponent of algebraic interaction decay and $d=3$ dimensionality in dipolar systems \cite{leitao_scalable_nodate}.

As a concrete example of this framework, consider a toy model of a Gaussian polarization distribution with spatial extent $R_*$.
\begin{align}
    \rho(\bm r) &= \frac{\rho_0}{\left(2\pi R_* \right)^{3/2}} \exp{-\frac{r^2}{2R_*^2}},
\end{align}
which is normalized such that $\int d\bm r \rho(\bm r)=\rho_0$. The Fourier transform is
\begin{align}
    \rho_{\bm k} &= \exp{-\frac{R_*^2}{2}k^2},
\end{align}
and the exchange kernel is thus
\begin{align}
    \mathcal{K}_{\bm Q}(\bm q) &= e^{-q^2 R_*^2}-\frac{1}{2}\left( e^{-R_*^2\left( \bm q+\bm Q\right)^2}+e^{-R_*^2\left( \bm q-\bm Q\right)^2}\right) \notag\\
    &= e^{-q^2 R_*^2}-e^{-q^2 R_*^2} e^{-Q^2 R_*^2} \cosh{2R_*^2\left(\bm q\cdot \bm Q \right)}.
\end{align}
We next perform the convolution by integrating over $\bm q$, 
\begin{align}
    \chi_{\bm Q} &= \rho_0 e^{-Q^2R_*^2} \int d\bm q \, A_{\hat{\bm \eta}}(\hat{\bm q}) \mathcal{V}(q/\Lambda)\, e^{-q^2R_*^2} \, \cosh{\left(2R_*^2qQ \left(\hat{q}\cdot\hat{Q} \right) \right)} \notag\\
    &= 4\pi\rho_0 A_{\hat{\bm \eta}}(\hat{\bm Q}) \, e^{-Q^2R_*^2} \, \int_0^\infty q^2 dq \, \mathcal{V}(q/\Lambda) \, e^{-q^2 R_*^2} G(2iqQR_*^2) \notag\\
    &= 4\pi \rho_0 A_{\hat{\bm \eta}}(\hat{\bm Q}) \, e^{-Q^2 R_*^2} \, \frac{1}{R_*^3}\int_0^\infty x^2 dx \, \mathcal{V}(x/\Lambda R_*) \, e^{-x^2} G(2iQR_* x) \notag\\
    &= \frac{4\pi}{3}\frac{\rho_0}{R_*^3} A_{\hat{\eta}}(\hat{Q}) F(QR_*,\left(\Lambda R_* \right)^{-1}),
\end{align}
where we performed a change of variables $x=qR_*$ before evaluating the remaining radial integral via an analytic series in the small parameter $\lambda \equiv \left(\Lambda R_* \right)^{-1}$. We obtain
\begin{align}
    F(q,\lambda) &= e^{-q^2}\int_0^\infty x^2dx \mathcal{V}(\lambda x) e^{-x^2}G(2iqx) \notag\\
    &= \sum_{n=0}^\infty \frac{6(n+1)}{(2n+3)!}\left( -1\right)^n F_n(q) \, \lambda^{2n}
\end{align}
where 
\begin{align}
    F_n(q) &= \int_0^\infty dx x^{2(n+1)} e^{-x^2} G(2iqx) \\
    &= \begin{cases}
        \frac{2\sqrt{\pi}q(2q^2-3)+3\pi e^{-q^2}\text{erfi}(q)}{16 q^3} & n=0 \\
        \begin{aligned}
        \frac{e^{-q^2}}{8q^2}\Gamma(n{+}\tfrac{1}{2}) \Big(& (2q^2{-}3)\, _1F_1(n{+}\tfrac{1}{2};\tfrac{1}{2};q^2) \\
        & + (4nq^2{+}3)\, _1F_1(n{+}\tfrac{1}{2};\tfrac{3}{2};q^2) \Big)
        \end{aligned} & n>0
    \end{cases}
\end{align}
is related to the imaginary error function $\text{erfi}(x)$ and the confluent hypergeometric function $_1F_1(a;b;x)$. 
To further theoretically quantify the sensitivity of the exchange field to the microscopic extent of the polarization observed in the main text, we introduce the notion of the \textit{spin exchange moment} \cite{mermin_absence_1966, lange_interaction_1967, halperin_hohenbergmerminwagner_2019}
\begin{align} \label{eq: exchange-moment}
    \kappa &= \langle\langle r^2\rangle \rangle \lbrack \rho \rbrack \notag\\
    &\equiv \int d\bm r w(\bm r) \rho(\bm r) \int d\bm r'\rho(\bm r') (r-r')^2 J(\bm r-\bm r') \notag\\
    &= \partial_{Q}^2 \chi_Q \bigg \vert_{Q=0}.
\end{align}
In the short-range Heisenberg model, the exchange moment is a finite, intensive quantity which guarantees the existence of gapless excitations above the ground state which spontaneously breaks the SU(2) symmetry. However, for the dipolar couplings relevant to our experiment, the exchange moment will diverge quadratically with system size. 

Numerically evaluating the exchange moment for the polarization distribution generated by optical pumping, Eq. \ref{eq: polarization-model}, we plot the result in Fig.~\ref{fig: SI10}(d). We observe a striking sensitivity of the spin exchange moment to the optical pumping times probed in experiment, which are demarcated by vertical dashed lines.  Critically, this analysis validates the claim that the exchange field is sensitive to the microscopic extent,  $Q_* \sim 1/R_*$, asserted in the main text. Indeed, sensitivity of the exchange field to both UV and IR information in three-dimensions is a hallmark of dipolar interactions, without an analog in short-range ferromagnets \cite{leitao_scalable_nodate}.
\\
\\
\subsubsection{Precession dynamics as a probe of collective behavior.}
In what follows, we consider the coherent dynamics of the collective coherence 
\begin{align}
    i \partial_t \langle S^+(t)\rangle_{\theta} &= -\langle \lbrack H, S^+(t) \rbrack \rangle
\end{align}
under three models of unitary many-body dynamics in the spin coherent state,
\begin{align}
    \rho_\theta = \prod_j \frac{1+P(\cos{\theta}X_j + \sin{\theta} Z_j)}{2}.
\end{align}
First, we analyze the dynamics in the case of the \textit{one-axis-twisting model}, which is collective for the trivial reason of being all-to-all coupled. 
\begin{align}
    H &= \frac{\chi}{N} S_z^2.
\end{align}
Computing the equation of motion for the collective spin coherence, we obtain
\begin{align}
    i \partial_t S^+ &= \lbrack S^+ , H \rbrack \notag\\
    &= -\frac{\chi}{N}\left( 2S_z+1  \right)S^+ \notag\\
    &\equiv  -\omega(S_z)S^+
\end{align}
such that 
\begin{align}
    S^+(t) &= \exp{\left(i \omega(S_z)t\right)} S^+.
\end{align}
\begin{align}
    \langle S^+(t) \rangle &= \langle e^{+i\chi t(2S_z+1)/N}S^+ \rangle \notag\\
    &= \sin{\theta} \left(\mathcal{G}_\theta(\chi t/N)\right)^{N-1} 
\end{align}
where $\mathcal{G}_\theta(\phi)$ is the single spin-$1/2$ generating function. This is easily computed 
    \begin{align}
        \mathcal{G}_\theta(\phi) &= \langle e^{-i \phi S_z} \rangle_{\theta} \notag\\
        &= \cos{\left(\frac{\phi}{2}\right)}-i P \cos{\theta} \sin{\left(\frac{\phi}{2}\right)}
    \end{align}
    for generic polarization of the spin,  $P$. Simplifying the result in the limit of large $N$, 
\begin{align}
    \langle S^+(t) \rangle &= \sin\theta \left( \cos{\left(\chi t/N \right)}+i\cos{\theta} \, \sin{\left(\chi t/N \right)}\right)^{N-1} \notag\\
    &= \sin \theta \left(1-\frac{\left(\chi t\right)^2/2N-i\cos{\theta} \, \chi t}{N} +\mathcal{O}\left(\left(\chi t/N\right)^3 \right)\right)^{N-1} \notag\\ 
    &\to \sin \theta \exp{\left(-\frac{\chi^2t^2}{2N}\right)} \exp{\left(i \cos{\theta}\, \chi t\right)}
\end{align}
where we assume $\chi t\ll N$. In summary, we have
\begin{align}
    \langle S^+(\chi t) \rangle_{\theta} &= \sin{\theta}\exp{\left(-\left( \frac{\left(\chi t/\mathcal{Q}\right)^2}{2}-i \cos{\theta} \, \chi t \right) \right)}  
\end{align}
where a \textit{twisting quality factor} $\mathcal{Q}\equiv \sqrt{N}$ is naturally identified.

We next consider the dynamics of a \textit{dilute dipolar Ising model}, serving as a \textit{local} version of the canonical collective nonlinearity discussed above. In particular, we consider
\begin{align}
    H &= \sum_{i<j} J(\bm r_{ij}) S_i^z S_j^z \notag\\
    &= \frac{1}{2} \sum_{ij} J(\bm r_{ij}) S_i^z S_j^z.
\end{align}
For this Hamiltonian, the local coherence operators evolve as 
\begin{align}
    i \partial_t S_j^+ &= \lbrack S_j^+, H \rbrack \notag\\
    &= -\omega_j^z S_j^+
\end{align}
where 
\begin{align}
    \omega_j^z &= \sum_{k} J(\bm r_{jk}) S_k^z
\end{align}
is the local precession frequency. Crucially, it is a static operator so the Heisenberg equation of motion can be integrated exactly to give
\begin{align}
    S_j^+(t) &= \exp{\left(+i \omega_j^z \,t\right)} S_j^+.
\end{align}
Computing the disorder-averaged twisting signal \cite{feldman_configurational_1996} gives,
    \begin{align}
        \overline{\langle S^+ \rangle_{\theta}} &= 
        \sin\theta \, \overline{\sum_j \langle e^{i \omega_j^z t} \rangle } \notag\\
        &=  \sin\theta \, \overline{\sum_j \prod_{k \neq j}\langle e^{i J(\bm r_{jk}) t S_k^z} \rangle } \notag\\ 
        &= \sin\theta \overline{\sum_j \prod_{k \neq j} \mathcal{G}_\theta(J(\bm r_{jk})t)} \notag\\
        &\approx  \sin\theta \left(1-\overline{(1-\mathcal{G}_\theta(J(\bm r)t)}\right)^{N-1} \notag\\
        &=  \sin\theta \left(1-\frac{n}{N}\int d\bm r \left(1-\mathcal{G}_\theta(J(\bm r)t) \right) \right)^{N-1} \notag\\
        &\to \sin\theta \, \exp{\left( - 2\pi n \int_0^\infty r dr \, \left( 1-\mathcal{G}_\theta\left(\frac{J_0 t}{r^3}\right) \right)  \right)}
    \end{align}
where $n$ is the areal spin density and $\mathcal{G}_\theta(\phi)$ is the single spin generating function. 
    
Therefore, the twisting signal gives
\begin{align}
    \overline{\langle S^+ \rangle_\theta} &= \sin \theta \, \exp{\left(-\left( \left( \gamma t \right)^{2/3} - i P\cos\theta \left( \chi t\right)^{2/3} \right)\right)}
\end{align}
and 
\begin{align}
    \gamma(\nu)  &= \frac{J_0 n^{1/\nu}}{2\times  3^{1/\nu}} \, \left( 2\pi \int \frac{du}{u^{1+\nu}} \sin{u} \right)^{1/\nu}; \\
    \chi(\nu)  &= \frac{J_0 n^{1/\nu}}{2\times  3^{1/\nu}} \, \left( 2\pi \int \frac{du}{u^{1+\nu}} \left(1-\cos{u} \right) \right)^{1/\nu}, 
\end{align}
where $\nu=2/3$ is the stretching exponent of free-induction decay. The quality factor is thus
\begin{align}
    \mathcal{Q} &\equiv \left(\frac{\chi}{\gamma}\right)^\nu \\
    &= \frac{\int_0^\infty \frac{du}{u^{1+\nu}} \sin{u}}{\int_0^\infty \frac{du}{u^{1+\nu}}(1-\cos{u}) } \notag \\
    &= \tan{\left(\frac{\pi\nu}{2} \right)}.
\end{align}
For two-dimensional dipoles, we thus have $\chi/\gamma= \tan{\pi/3} = \sqrt{3}$. In the main text, we further multiply this number by $\cos{\pi/4}=1/\sqrt{2}$ to compare our measurements to this result. 
\\
Optimizing the twisting signal, we obtain
\begin{align}
    \Im \overline{\langle S^+ \rangle_\theta } &= \sin\theta \sin{\left(P\cos{\theta}(\chi t)^{2/3}\right)} e^{-(\gamma t)^{2/3}} \\
    \implies\partial_t \Im \overline{\langle S^+ \rangle_\theta } &= \sin{\theta} \left(\frac{2\chi}{3}P\cos{\theta} (\chi t)^{-1/3} \cos{\phi}e^{-(\gamma t)^{2/3}} - \sin{\phi} \frac{2 \gamma}{3}(\gamma t)^{-1/3} e^{-(\gamma t)^{2/3}} \right) \notag \\
    \implies \tan\phi_* &= P(\chi/\gamma)^{2/3} \cos\theta \equiv P \mathcal{Q} \cos{\theta}
\end{align}

where $\phi_*=\chi t_*$ is the optimal twisting angle.

Furthermore, we can analytically obtain the optimal twisting signal
\begin{align}
    \Im \overline{\langle S^+(t_*)\rangle_\theta } &= \sin\theta \cos\theta \, \frac{P \mathcal{Q}  }{\sqrt{\left(P \mathcal{Q} \cos\theta\right)^2+1}} \exp{-\frac{\arctan{P\mathcal{Q}\cos{\theta}}}{P\mathcal{Q}\cos\theta}},
\end{align}
 whose asymmetry under $\theta \to \pi/4-\theta$ is controlled directly by $\mathcal{Q}$, as alluded to in the main text. 

Finally, we now argue for the form of the precession signal for the \textit{dipolar Heisenberg model with a spin spiral initial state}. Here we thus have
\begin{align}   
    \langle S^+_{\bm Q}(t)\rangle_\theta &= \text{Tr}\left(V_{\bm Q}^\dagger S^+ V_{\bm Q} e^{-i H_0 t}V_{\bm Q}^\dagger \rho_{\theta} V_{\bm Q}  e^{+i H_0 t}\right),
\end{align}
which can be analyzed in the same framework as above, so long as we perform a local transformation on the underlying Heisenberg Hamiltonian
\begin{align}
    i \partial_t\langle S^+_{\bm Q}(t) \rangle &= \langle \lbrack H_{\bm Q}, S_+\rbrack \rangle_{\rho_\theta}.   
\end{align}
While the resulting equation of motion is no longer integrable like the examples above, the early-time expansion above can still be employed to probe the solution \cite{leitao_scalable_nodate}. In particular, the early-time expansion gives a very similar expression for the twisting amplitude
\begin{align}
    \langle S^+_{\bm Q}(t)\rangle &\approx \sin{\theta} \exp{\left( -\left(\sin^2\theta \, \gamma_{\bm Q}^2t^2-i \chi_{\bm Q} \cos{\theta} \, t \right) \right)}
\end{align}
where 
\begin{align}
    \gamma_{\bm Q}^2 &= \sum_{j} J^2(\bm r_j)(1-\cos{\bm Q \cdot \bm r_j})^2 
\end{align}
is the intrinsic early-time spiral decay rate at the $SU(2)$ point. To get intuition for this term, note that under the assumption of emergent hydrodynamics, the infinite temperature spin autocorrelation function 

\begin{align}
    C_{\bm Q}(t)=\frac{\text{Tr}\left( S^+_{\bm Q}(t) S^-_{-\bm Q}(0)\right) }{\text{Tr}\left( S^+_{\bm Q}(0) S^-_{-\bm Q}(0)\right) }
\end{align}
should accurately describe the dynamics of spin spirals in our system. Expanding this autocorrelator at early-times gives us the term:
\begin{align}
    C_{\bm Q}(t) &= 1-\gamma_{\bm Q}^2 t^2 + \mathcal{O}(t^4),\\
   \gamma_{\bm Q}^2 &=  \frac{\text{Tr}\left(\lbrack H_{\bm Q}, S^+ \rbrack \lbrack H_{\bm Q}, S^- \rbrack  \right)}{\text{Tr}\left(S^+ S^-  \right)},
\end{align}
which is nothing but the second moment of the mean-field exchange operator in the Heisenberg picture. In other words, while the exchange field strength in the polarized state is given by $\chi_{\bm Q}$  that is a coherent weighted average of the spiral texture, the exchange field noise $\gamma_{\bm Q}$ is determined by an incoherent (root-mean-square) average of the spiral texture.  

For three-dimensional dipolar systems, this \textit{incoherent} average over the spiral texture can be shown to scale as 
\begin{align}
    \gamma_{\bm Q}^2 &= n\int d\bm r   \frac{A^2_{\eta}(\hat{\bm r})}{r^6} (1-\cos{\left(\bm Q \cdot \bm r \right)})^2 \notag\\
    &= n \int_0^\infty r^3 \frac{dr}{r} \frac{1}{r^6} \int d\hat{\bm r} \, A_{\hat{\bm \eta}}^2(\hat{\bm r}) \, \left(1-\cos{\left(Qr \, \hat{\bm Q} \cdot \hat{\bm r} \right)} \right)^2 \notag\\
    &\equiv n \int_0^\infty r^3 \frac{dr}{r} \frac{1}{r^6} F_{\hat{Q},\hat{\eta}}(Qr) \notag\\
    &= n Q^3\int_0^\infty \frac{du}{u^4}F_{\hat{Q},\hat{\eta}}(u)
\end{align}
as $Q \to 0$, which can be easily seen via the rescaling $u=Qr$. Crucially, this implies that the quality factor relevant to spiral twisting scales as $\chi_{\bm Q}/\gamma_{\bm Q}\sim Q^{-3/2}\sim N^{1/2}$ when $Q\to Q_*\equiv N^{-1/3}$, as asserted in the main text.

\clearpage

\bibliography{NonlinearDynamics}